\documentclass[12pt,a4paper]{article}
\usepackage{preamble}

\begin{document}

\setlength{\abovedisplayskip}{3pt}
\setlength{\belowdisplayskip}{3pt}

\pagestyle{plain}

\makeatletter
\@addtoreset{equation}{section}
\makeatother
\renewcommand{\theequation}{\thesection.\arabic{equation}}

	\pagestyle{empty}
	\vspace{0.5cm}

    \begin{center}
    	{\LARGE \bf{Further Bounding the Kreuzer-Skarke Landscape} \\[10mm]}
	\end{center}
	
	\begin{center}
		\scalebox{0.95}[0.95]{{\fontsize{14}{30}\selectfont Nate MacFadden \orcidlink{0000-0002-8481-3724},$^{a}$ Stepan Yu. Orevkov,$^{b}$ and Michael Stepniczka \orcidlink{0000-0002-3838-6653}$^{a}$}} 
	\end{center}

	\begin{center}
		\textsl{$^{a}$Department of Physics, Cornell University, Ithaca, New York 14853 USA}\\
        \textsl{$^{b}$Institut de Math{\'e}matiques de Toulouse, Universit{\'e} de Toulouse,
Toulouse 13062 France}\\
        \vspace{.35cm}
        \href{mailto:njm222@cornell.edu}{\hspace{3.8pt}nate.macfadden@gmail.com},
        \href{mailto:stepan.orevkov@math.univ-toulouse.fr}{\hspace{3.8pt}stepan.orevkov@math.univ-toulouse.fr}, \href{mailto:ms3296@cornell.edu}{\hspace{3.8pt}ms3296@cornell.edu}
		
		\vspace{1cm}
		\normalsize{\bf Abstract} \\[8mm]
	\end{center}


Batyrev's construction provides a map from fine, regular, star triangulations (FRSTs) of $4$D reflexive polytopes to smooth Calabi-Yau threefolds (CYs). We prove that there are at most $10^{296}$ diffeomorphism classes of CYs produced in this manner, improving \cite{Demirtas:2020dbm}'s upper bound of $10^{428}$. 
To show this, we make use of the fact that any two FRSTs with the same $2$-face restrictions give rise to diffeomorphic CYs and bound the number of such `$2$-face equivalence classes' for all polytopes with Hodge number $h^{1,1} \geq 300$. 
We also put a lower bound of $10^{276}$ on the number of $2$-face equivalence classes, but emphasize that this is not a lower bound on the number of diffeomorphism classes of CYs, as distinct $2$-face equivalence classes may give rise to diffeomorphic threefolds. 
\newpage

\setcounter{page}{1}
\pagestyle{plain}
\renewcommand{\thefootnote}{\arabic{footnote}}
\setcounter{footnote}{0}
%
%
\hypersetup{linkcolor=black}
\tableofcontents
\newpage

\section{Introduction}\label{section: intro}
This paper concerns the number of distinct $4$D effective field theories defined as compactifications of string theory over smooth Calabi-Yau threefolds (CYs). Since non-diffeomorphic threefolds give rise to physically inequivalent effective field theories, one is interested in the count of diffeomorphism classes of CYs. Precisely counting the number of such classes is an active area of research, with current methods only applying to a small subset of known CYs\cite{gendler2023counting,oxford}. Rather than furthering this precise count, we follow \cite{Demirtas:2020dbm} and substantially tighten the best-known \textit{bounds} on the number of diffeomorphism classes of CYs that are producible via Batyrev's construction\cite{Batyrev:1993oya}. 

The motivation for studying such CYs is that they are both the most numerous across any known construction and they are remarkably accessible. The accessibility is due to the combinatorial nature of Batyrev's construction, which, for our purposes, defines a smooth Calabi-Yau threefold from the data of a triangulation (satisfying various properties -- see \Cref{subsec:frts_wall}) of any $4$D reflexive polytope. Recall that a polytope is called \textit{reflexive} if both it and its polar dual are lattice. All 473,800,776 such $4$D reflexive polytopes were enumerated by Kreuzer and Skarke in what is known as the Kreuzer-Skarke database \cite{Kreuzer:2000xy}, and can be retrieved on-demand. Triangulations are well suited for computer-based studies, so these CYs can be understood efficiently via the use of combinatorial, triangulation-focused algorithms. For example, using the software package \cytools\cite{Demirtas:2022hqf}, researchers routinely study millions of such CYs in the scope of a paper. Together, then, the abundance and accessibility of these CYs make them one of the most exciting frameworks for constructing effective theories of quantum gravity -- see \cite{McAllister:2024lnt, Demirtas:2021nlu,sheridan2024fuzzyaxionsassociatedrelics} for some example studies.

Upper bounds on the number of diffeomorphism classes of CYs producible via Batyrev's construction were originally obtained in \cite{Demirtas:2020dbm}, whereby bounding the number of relevant triangulations, it was demonstrated that there are at most $10^{928}$ such classes. In the same paper, the authors then made use of Wall's theorem\cite{wall} to substantially tighten the upper bound by 500 orders of magnitude to $10^{428}$. 
This theorem, as further discussed in \Cref{subsec:frts_wall}, implies that two such CYs are diffeomorphic if their underlying triangulations have identical $2$-face restrictions. In this way, one defines equivalence classes of triangulations which we call `$2$-face equivalence classes'. Clearly, then,
\begin{equation}
\label{eq:wall}
    \textrm{\# diffeomorphism classes of CYs} \leq \textrm{\# $2$-face equivalence classes}.
\end{equation}
To obtain the improved upper bound of $10^{428}$, the authors of \cite{Demirtas:2020dbm} bounded the number of $2$-face equivalence classes from above. Prior to the present work, it was unknown how loose this upper bound of $2$-face equivalence classes was; in this work, we strengthen it to $10^{296}$, and furthermore, show that it cannot drop beneath $10^{276}$. We nevertheless stress: a lower bound on the number of $2$-face equivalence classes is \textit{not} a lower bound on the number of diffeomorphism classes of CYs, as two CYs with distinct $2$-face restrictions can be diffeomorphic\footnote{In fact, to the best of the authors' knowledge, a non-trivial lower bound on the number of diffeomorphism classes of CYs arising from Batyrev's construction is not known, beyond direct enumeration at small Hodge number $h^{1,1}$.}.

Out of the $\sim$half-billion $4$D reflexive polytopes, only a handful of them support a non-negligible count of 2-face equivalence classes (and hence, only this handful can potentially support a non-negligible count of CYs). In particular, the best prior bound on the number of $2$-face equivalence classes depends (roughly) exponentially on the Hodge number $h^{1,1}$. This is most apparent when comparing the unique polytope with the maximal value $h^{1,1}=491$,
\begin{equation}\label{eq:491}
    \Delta^\circ_{491} \coloneqq \text{conv}\begin{pmatrix}
        -63 & 0 & 0 & 1 & 21\\
        -56 & 0 & 1 & 0 & 28\\
        -48 & 1 & 0 & 0 & 36\\
        -42 & 0 & 0 & 0 & 42
    \end{pmatrix},
\end{equation}
to all others ($h^{1,1}\leq 462$) in \cite{Demirtas:2020dbm}'s bounds: while $\Delta^\circ_{491}$ supports at most $10^{428}$ $2$-face equivalence classes, all other polytopes together support only at most $10^{402}$ $2$-face equivalence classes. To improve bounds on the count of distinct CYs, then, it suffices to limit attention to the largest $h^{1,1}$ polytopes. It is for this reason that we study all 4D reflexive polytopes with $h^{1,1} \geq 300$ in this paper, although we note that the methods utilized in this work are relevant to any polytope. In particular, we will use $\Delta^\circ_{491}$ as a running example.

To build up to our improved bound of $10^{296}$
, we first recall Batyrev's construction, Wall's theorem, and $2$-face equivalence classes in \Cref{subsec:frts_wall}. With this background covered, we then demonstrate in \Cref{section: upper bounds} how to reduce the upper bound on the number of distinct CYs. 
Finally, we demonstrate a lower-bound of such $2$-face equivalence classes in \Cref{section: lower bounds} that is only $\sim20$ orders of magnitude from the upper bound. We conclude in \Cref{section: conclusion}.

\subsection{Review of FRSTs and Wall's theorem} \label{subsec:frts_wall} 

To understand Batyrev's construction\cite{Batyrev:1993oya} to the level necessary to count the corresponding CYs, we first must review triangulations. Let $P$ be a convex lattice polytope and, following \cite{Demirtas:2020dbm}, let $F_P(d)$ denote its collection of $d$-dimensional faces. 
By a `triangulation' of $P$, we mean\footnote{This definition can actually be generalized -- see e.g. \cite{vex,Berglund_2018}.} a simplicial complex subdividing $P$ with vertices taken from $P\cap\mathbb{Z}^4$. 

The polytopes of interest for Batyrev's construction are $4$D reflexive polytopes -- denote them as $\Delta^\circ$. While one can use all points $\Delta^\circ\cap\mathbb{Z}^4$ in their triangulations, it is beneficial to further restrict vertices to
\begin{equation}
    \mathbf{A}(\Delta^\circ) \coloneqq \{ p\in\Delta^\circ\cap\mathbb{Z}^4 : p\not\in\text{relint}(f) \text{ for all } f\in F_{\Delta^\circ}(3) \}
\end{equation}
for such reflexive polytopes. In other words, one discards any lattice points strictly interior to facets. This restriction is useful because (see, e.g., Appendix B of \cite{2017hodge}) points strictly interior to facets of $\Delta^\circ$ do not affect the resulting CY and because every triangulation $\mathcal{T}'$ defined over $\Delta^\circ\cap\mathbb{Z}^4$ has a `$2$-face equivalent' triangulation $\mathcal{T}$ defined over $\mathbf{A}(\Delta^\circ)$\cite{Demirtas:2020dbm}. See \cite{macfadden2023efficient} for such a map $\mathcal{T}'\to\mathcal{T}$. Thus, no diffeomorphism classes are missed by such a restriction. The restriction can be thought of as simplifying the polytope by discarding `irrelevant' points. 

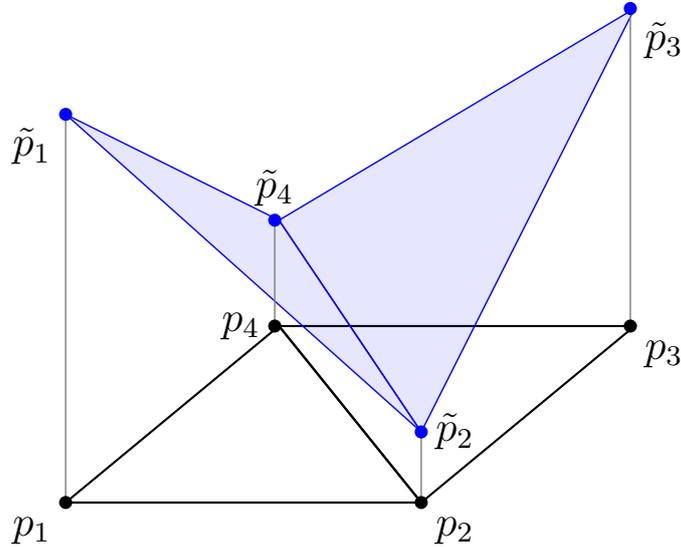
\begin{figure}[ht]
    \centering
    \resizebox{0.6\textwidth}{!}{
    \begin{tikzpicture}[scale=3.5]
        \draw[black, semithick] (0, 0) -- (1, 0) -- (0+0.6, 0.5) -- cycle;
        \draw[black, semithick] (1+0.6, 0.5) -- (1, 0) -- (0+0.6, 0.5) -- cycle;
    
        \fill[blue,opacity=0.1] (0, 0+1.1) -- (1, 0+0.2) -- (0+0.6, 0.5+0.3) -- cycle;
        \fill[blue,opacity=0.1] (1+0.6, 0.5+0.9) -- (1, 0+0.2) -- (0+0.6, 0.5+0.3) -- cycle;
        \draw[blue] (0, 0+1.1) -- (1, 0+0.2) -- (0+0.6, 0.5+0.3) -- cycle;
        \draw[blue] (1+0.6, 0.5+0.9) -- (1, 0+0.2) -- (0+0.6, 0.5+0.3) -- cycle;
    
        \foreach \x in {0,1} {
            \foreach \y in {0,0.5} {
                \pgfmathsetmacro\height{
                    ifthenelse(\x==0 && \y==0, 1.1,
                    ifthenelse(\x==1 && \y==0, 0.2,
                    ifthenelse(\x==1 && \y==0.5, 0.9, 
                    ifthenelse(\x==0 && \y==0.5, 0.3, ))))
                }
                \draw[thin, gray] (\x+1.176*\y,\y) -- (\x+1.176*\y,\y+\height);
            }
        }
    
        \node[fill=black, circle, inner sep=1.3pt] at (0,0) {};
        \node[fill=black, circle, inner sep=1.3pt] at (1,0) {};
        \node[fill=black, circle, inner sep=1.3pt] at (1+1.176*0.5,0.5) {};
        \node[fill=black, circle, inner sep=1.3pt] at (0+1.176*0.5,0.5) {};

        \node[below left] at (0,0) {$p_1$};
        \node[below right] at (1,0) {$p_2$};
        \node[below right] at (1+1.176*0.5,0.5) {$p_3$};
        \node[left] at (0+1.176*0.5,0.5) {$p_4$};
    
        \node[fill=blue, circle, inner sep=1.3pt] at (0,0+1.1) {};
        \node[fill=blue, circle, inner sep=1.3pt] at (1,0+0.2) {};
        \node[fill=blue, circle, inner sep=1.3pt] at (1+1.176*0.5,0.5+0.9) {};
        \node[fill=blue, circle, inner sep=1.3pt] at (0+1.176*0.5,0.5+0.3) {};

        \node[below left] at (0,0+1.1) {$\tilde{p}_1$};
        \node[right] at (1,0+0.2) {$\tilde{p}_2$};
        \node[below right] at (1+1.176*0.5,0.5+0.9) {$\tilde{p}_3$};
        \node[above] at (0+1.176*0.5,0.5+0.3) {$\tilde{p}_4$};

    \end{tikzpicture}
    }
    \caption{Diagram of the `lifting' procedure defining regular triangulations. The points $p_1$, $p_2$, $p_3$, and $p_4$ are embedded into $\mathbb{R}^3$ and then lifted by heights $\omega_1=1.1$, $\omega_2=0.2$, $\omega_3=0.9$, and $\omega_4=0.3$. The convex hull of the lifted point configuration is a $3$-simplex whose lower faces are plotted in blue. Projecting out the lifted coordinate generates the regular triangulation plotted in black. Figure modified from \cite{macfadden2023efficient}.}
    \label{figure: regularity}
\end{figure}

Label the points $\mathbf{A} = \{p_1, \dots, p_N\}$. For a triangulation $\mathcal{T}$ of $\Delta^\circ$ to be suitable for Batyrev's construction, it must also satisfy the following three properties:
\begin{enumerate}
    \item $\mathcal{T}$ is \textit{fine}, i.e. every point $p\in\mathbf{A}$ is the vertex of at least one simplex\footnote{N.B.: for 2D lattice polygons, `fine' is synonymous with `unimodular' or `primitive'.} $\sigma\in\mathcal{T}$,
    \item $\mathcal{T}$ is \textit{regular}, i.e. it can be obtained as the lower faces of $\textrm{conv}\left(\{(p_i,\omega_i): 1\leq i\leq |\mathbf{A}|\}\right)$ for some `heights' $\omega=(\omega_1,\dots,\omega_N)$ (see \Cref{figure: regularity}), and 
    \item $\mathcal{T}$ is \textit{star}, i.e. the unique lattice point $p\in\text{int}(\Delta^\circ)$ (typically taken to be the origin) is a vertex of every 4-simplex $\sigma\in\mathcal{T}$.
\end{enumerate}
If $\mathcal{T}$ is fine, regular, and star, we call it an `FRST'; if $\mathcal{T}$ only obeys a subset of the properties, then only the corresponding letters are kept. Note, these properties can directly be generalized to any convex polytope $P$, with the caveat that the `star' property only makes sense if $P$ has a unique interior point.

With this terminology, we now recall Batyrev's construction\cite{Batyrev:1993oya}. Consider an FRST $\mathcal{T}$ of a $4$D reflexive polytope $\Delta^\circ$. Since $\mathcal{T}$ is star, it defines a complete simplicial fan and hence a toric variety. Additionally, since $\mathcal{T}$ is regular, this toric variety is projective and hence K{\"a}hler. Batyrev's construction, then, is the identification of (the closure of) a generic anticanonical hypersurface of this toric variety as a (potentially singular) Calabi-Yau threefold. This hypersurface is given by $F=0$, where the polynomial $F$ is implicitly defined to have a Newton polytope dual to $\Delta^\circ$. 
In fact, since $\mathcal{T}$ is also fine, the singularities of the toric variety are sufficiently mild such that the CY is smooth\footnote{By \cite{Batyrev:1993oya}, this smoothness relies on the dimension of the reflexive polytope being $\leq 4$; otherwise, this procedure may not resolve all singularities intersecting a generic Calabi-Yau hypersurface.}. 
Since there are $\leq 10^{928}$ FRSTs of $4$D reflexive polytopes\cite{Demirtas:2020dbm}, there are na\"ively $\leq 10^{928}$ such CYs.

For example, the dual polytope of $\Delta_{491}^\circ$ is 
\begin{equation*}
    \text{\rm conv\,}\left(\begin{matrix}
     -1 &  -1 &  -1 & -1 &  1 \\
     -1 &  -1 & -1 &  2 &  -1 \\
     -1 & -1 &  6 &  -1 &  -1 \\
    2 &  4 &  -4 & 0 & 1 \end{matrix}\right),
\end{equation*}
and thus the associated CYs are closures of generic hypersurfaces of the form
\begin{equation*}
\begin{split}
  & \, c_0 
    + c_1\frac{y}x + c_2\frac{z}x + c_3\frac{z}y + c_4\frac{w}x +  
c_5\frac{w}y + c_6\frac{w}z
\\&\quad
    + c_7\frac{y^2}{xz} + c_8\frac{z^2}{xy} + + c_{9}\frac{z^2}{xw} 
    + c_{10}\frac{z^2}{yw}
    + c_{11}\frac{w^2}{xy} + c_{12}\frac{w^2}{xz} + c_{13}\frac{w^2}{yz}
\\&\quad
    + c_{14}\frac{xw}{yz} + c_{15}\frac{yz}{xw} + c_{16}\frac{yw}{xz} +
c_{17}\frac{zw}{xy}
    + c_{18}\frac{z^3}{xw^2}
\\&\quad
    + c_{19}\frac{z^3}{xyw}
    + c_{20}\frac{z^4}{xyw^2}
    + c_{21}\frac{z^5}{xyw^3}
    + c_{22}\frac{z^6}{xyw^4}
    + c_{23}\frac{w^2}{xyz} + c_{24}\frac{w^3}{xyz} +
c_{25}\frac{w^4}{xyz} = 0.
\end{split}
\end{equation*}

While the bound on FRSTs is directly a bound on the number of diffeomorphism classes of CYs, Wall's theorem\cite{wall} enabled \cite{Demirtas:2020dbm} to make stronger statements about when two FRSTs define equivalent CYs. Wall's theorem states that the diffeomorphism classes of such (simply connected, torsion-free) CYs\footnote{Note, in \cite{batyrev2005integral}, Batyrev and Kreuzer find 16 mirror pairs of $4$D reflexive polytopes which fail these conditions, meaning Wall's theorem does not apply. None of these polytopes appear in the collection considered in this work, i.e. those with $h^{1,1} \geq 300$.} are determined by their Hodge numbers, second Chern class, and triple intersection numbers. As is discussed in \cite{Demirtas:2020dbm}, the Hodge numbers are determined by the underlying polytope, while the second Chern class and intersection numbers are determined by the $2$-face restrictions of the triangulation. Thus, if two FRSTs $\mathcal{T}_1$ and $\mathcal{T}_2$ of the same polytope $\Delta^\circ$ have the same $2$-face restrictions, then their associated CYs are diffeomorphic. It is therefore convenient to define an equivalence relation such that $\mathcal{T}_1\sim\mathcal{T}_2$ if and only if $\mathcal{T}_1$ has the same $2$-face restrictions as $\mathcal{T}_2$ (i.e., if their $2$-skeletons have the same triangulations); denote this as `$2$-face equivalence'\footnote{One can also ask for $2$-face equivalence up to polytope automorphism. As $\Delta^\circ_{491}$ has the largest count of $2$-face equivalence classes but only has two automorphisms, we do not consider this.}. 

In this language, we are interested in the count of $2$-face equivalence classes coming from all polytopes in the Kreuzer-Skarke database. \cite{Demirtas:2020dbm} demonstrated that this quantity is at most $10^{428}$. The leading contribution comes from 2-face equivalence classes of $\Delta_{491}^\circ$; the rest of the Kreuzer-Skarke polytopes together contribute $10^{402}$ to this bound. In the present paper, we show that the number of 2-face equivalence classes of $\Delta_{491}^\circ$ is still the leading contribution, with a true count bounded between $10^{276}$ and $10^{296}$. Moreover, we show that the rest of the polytopes together contribute at most $10^{279}$ to this bound. 
We leave it to future work to apply this to all Kreuzer-Skarke polytopes.

\section{Upper Bounds}\label{section: upper bounds}

For a given polytope $\Delta^\circ$, each $2$-face equivalence class is defined by the fine, regular triangulations (i.e., FRTs) of the $2$-faces $f\in F_{\Delta^\circ}(2)$. However, as will be further discussed in \Cref{section: lower bounds}, not every collection of FRTs of 2-faces can be realized as simultaneously descending from any FRST. Thus, the number of $2$-face equivalence classes defined by $\Delta^\circ$ is bounded by the number of distinct ways to assign FRTs to these $2$-faces:
\begin{equation}\label{eq: upper strategy}
    \text{\#$2$-face equivalence classes of }\Delta^\circ \leq \prod_{f\in F_{\Delta^\circ}(2)} N_{\text{FRT}}(f).
\end{equation}
To bound the number of $2$-face equivalence classes, it then suffices to evaluate $N_{\text{FRT}}(f)$ for each $2$-face $f$ of $\Delta^\circ$. Unfortunately, it is generally computationally expensive to count the number of FRTs for a lattice polygon. 
Fortunately, most $2$-faces arising in the Kreuzer-Skarke database are sufficiently small for the cost to be negligible. For instance, despite having a moderately high $h^{1,1}=60$ and a sizable bound of at most $ 10^{36}$ $2$-face equivalence classes, the example in Section 4.2 of \cite{MacFadden:2024him} has $N_{\text{FRT}}(f)\leq 724$ for all of its $2$-faces. Said polytope generates this upper bound not via large $2$-faces, but instead via many $2$-faces each with a moderate number of FRTs.

In contrast, \cite{Demirtas:2020dbm}'s bound on $2$-face equivalence classes is dominated by $\Delta^\circ_{491}$ 
despite it having only ten $2$-faces, the minimum possible number of faces in $4$D. This is possible because three of its $2$-faces, $f_8$, $f_9$, and $f_{10}$ are considerably large and therefore admit many FRTs -- see \Cref{table: 2-faces}. These $2$-faces each have enough FRTs such that directly counting them is generally computationally infeasible, primarily due to checking regularity. Regularity can be checked by solving a linear feasibility problem $Hx > 0$ for $x$, where $H$ is a matrix determined by the simplices of each triangulation. Needless to say, it is computationally infeasible to solve e.g. $\geq10^{180}$ such problems.  Nevertheless, we \textit{are} able to determine the exact number of FRTs of $f_8$ because, as will be discussed in \Cref{section: lower bounds}, every fine triangulation of $f_8$ is also regular. We are not so fortunate with $f_9$ and $f_{10}$.

\begin{table}[!htp] 
\begin{center}
    \begin{tabular}{|p{0.1\linewidth}|p{0.3\linewidth}|p{0.2\linewidth}|p{0.23\linewidth}|}
        \hline
        2-face & affinely equivalent polygon & FRT lower bound & FRT upper bound \\
        \hline
        \hline
        $f_{1}$ & $\text{conv}\begin{pmatrix}
                    0 & 1 & 0\\
                    0 & 0 & 1\\
                \end{pmatrix}$ & $1$ & $1$ \\
        \hline
        $f_{2},f_{3}$ & $\text{conv}\begin{pmatrix}
                    0 & 2 & 0\\
                    0 & 0 & 3\\
                \end{pmatrix}$ & $5$ & $5$ \\
        \hline
        $f_{4},f_{5}$ & $\text{conv}\begin{pmatrix}
                    0 & 2 & 0\\
                    0 & 0 & 7\\
                \end{pmatrix}$ & $204$ & $204$ \\
        \hline
        $f_{6},f_{7}$ & $\text{conv}\begin{pmatrix}
                    0 & 3 & 0\\
                    0 & 0 & 7\\
                \end{pmatrix}$ & $19594$ & $19594$ \\
        \hline
        $f_{8}$ & $\text{conv}\begin{pmatrix}
                    0 & 2 & 0\\
                    0 & 0 & 84\\
                \end{pmatrix}$ & $2.02 \times 10^{35}$ & $2.02 \times 10^{35}$ \\
        \hline
        $f_{9}$ & $\text{conv}\begin{pmatrix}
                    0 & 3 & 0\\
                    0 & 0 & 84\\
                \end{pmatrix}$ & $8.73 \times 10^{61}$ & $7.61 \times 10^{65}$ \\
        \hline
        $f_{10}$ & $\text{conv}\begin{pmatrix}
                    0 & 7 & 0\\
                    0 & 0 & 84\\
                \end{pmatrix}$ & $3.90 \times 10^{167}$ & $1.96 \times 10^{180}$ \\
        \hline
    \end{tabular}
\end{center}
\caption{Affinely equivalent polygons to the 2-faces of $\Delta_{491}^\circ$. Both lower bounds and upper bounds on the number of fine regular triangulations (FRTs) of each 2-face are cited.}
\label{table: 2-faces}
\end{table}

\begin{figure}[ht]
    \centering
    \includegraphics[width=\textwidth]{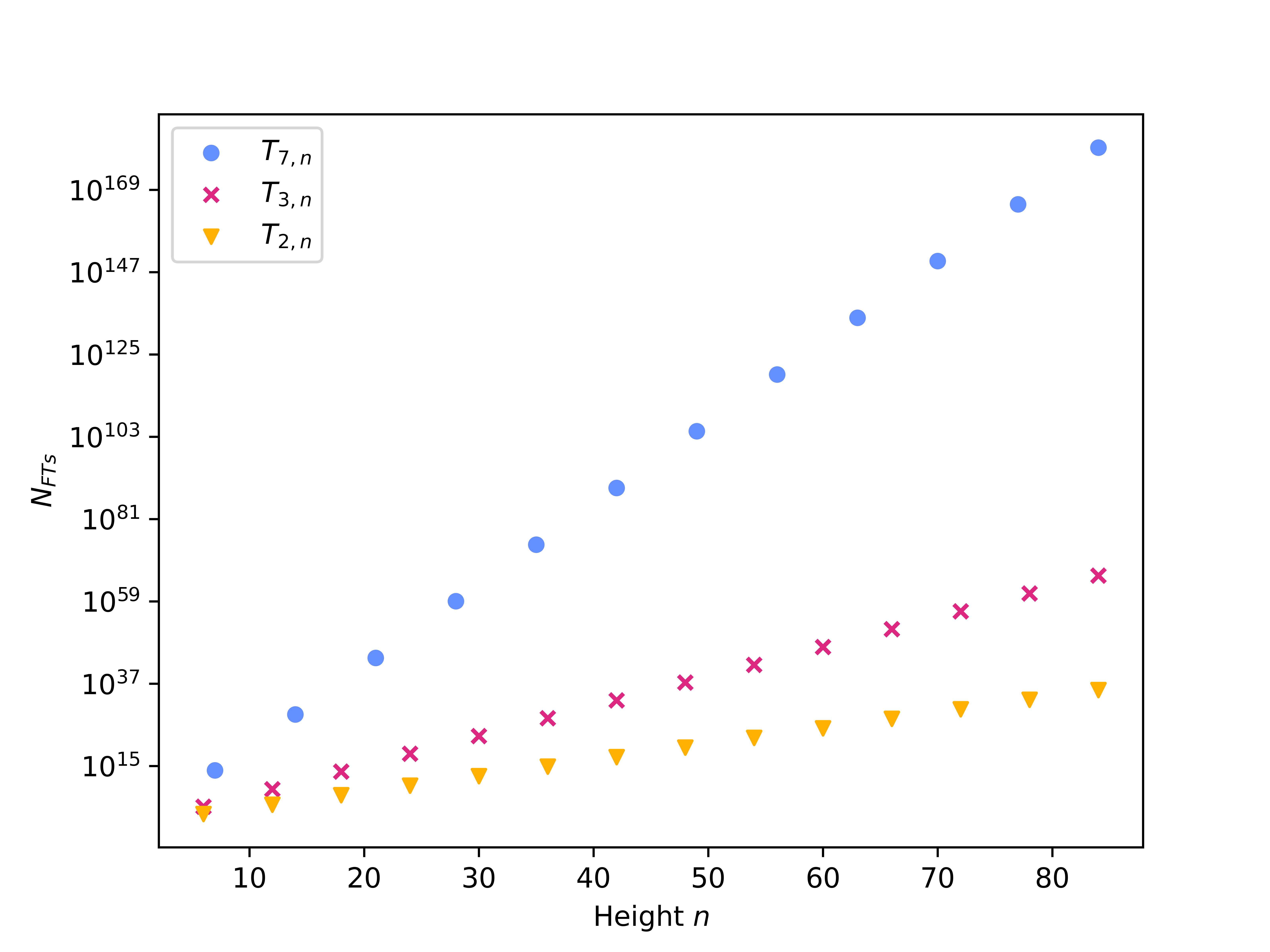}  
    \caption{The number of fine triangulations of right lattice triangles $T_{2,n}, T_{3,n},$ and $T_{7,n}$ as a function of the height $n$. In particular, $T_{2,84}$, $T_{3,84}$, and $T_{7,84}$ correspond to $f_8$, $f_9$, and $f_{10}$ respectively.}
    \label{figure: capacities}
\end{figure}

While it is desirable to use counts $N_{\text{FRT}}$ in (\ref{eq: upper strategy}), we can (and oftentimes do) also use upper bounds of these counts.
This was the exact strategy used in \cite{Demirtas:2020dbm} to bound $N_\textrm{CY}\leq 10^{428}$, for which the authors used exact counts $N_{\text{FRT}}(f)$ for faces $f_1, \dots, f_7$ and Anclin's upper bound\footnote{A stronger bound for lattice rectangles exists \cite{Welzl,WelzlSlides} which can be adapted to other lattice polygons.}\cite{anclin2003} on $N_\FT$ for $f_8$, $f_9$, and $f_{10}$. As there is no practical algorithm for determining $N_{\text{FRT}}$ for sufficiently large polygons, we instead make improvements by utilizing the general methods of \cite{Kaibel_Ziegler_2003,orevkov2022counting} to evaluate $N_{\text{FT}}$ for all $2$-faces \textit{exactly}\footnote{We note that in \cite{Demirtas:2020dbm}, the authors used $N_{\text{FRT}}$ rather than $N_{\text{FT}}$ for $2$-faces with at most 17 lattice points. While we include these data when relevant in \Cref{appendix: data}, to the precision reported in our bounds below, these differences do not matter. Therefore, should one wish, one may replace the inequality (\ref{eq: upper strategy}) with $N_{\text{FT}}(f)$ for all 2-faces $f$.}.
See \Cref{appendix: method,appendix: complexity} for more details regarding the algorithm.

Since the counts of $N_{\text{FT}}$ become large, we will want to take care in the size of the integers that we store. To address this, we use the Chinese Remainder Theorem. Specifically, our procedure for evaluating $N_{\mathrm{FT}}$ for any $2$-face is:
\begin{enumerate}
    \item select the largest $N$ prime numbers $\leq 32767$ such that their product is larger than Anclin's bound on $N_{\mathrm{FT}}$, thus enabling use of the CRT,
    \item evaluate, using the methods of \cite{orevkov2022counting}, $N_{\mathrm{FT}}\mod p$ for each prime number, and then
    \item combine the computations using the CRT to evaluate $N_{\mathrm{FT}}$.
\end{enumerate}
Our selection of primes ensures that all counts appearing in the computation 
can be stored as $16$-bit integers\footnote{In hindsight, we note that \textit{unsigned} 16-bit integers allow one to use primes up to $2^{16}-1=65535$.
}. This measure is necessary because, even with just $16$-bit integers, each computation required a significant $\mathcal{O}$(40GB) RAM.  Note that since the time of performing the initial computations, the \texttt{C} code has been improved to use 32-bit offsets rather than 64-bit pointers and has been parallelized \cite{orevkov2025asymptotics,orevkov_webpage}. The memory usage can be further optimized if we restrict our focus to a specific geometry, e.g. to right lattice triangles. However, while this would suffice for studying $\Delta^\circ_{491}$, it would not suffice for studying the set of $2$-faces arising from all Kreuzer-Skarke polytopes with $h^{1,1} \geq 300$.

\subsection{Example: upper bounds for $\Delta^\circ_{491}$}\label{subsection: upper results} 
We now compute the precise number of fine triangulations for the large 2-faces of $\Delta^\circ_{491}$. Take, for example, $f_{10}$. This $2$-face had the largest value for Anclin's bound, $N_{\mathrm{FT}}(f_{10})< 10^{251}$. Thus, the following $n=57$ primes were used to obtain the correct result.
\begin{equation}
    10^{251} \leq \prod_{i=1}^{n} p_i = 29443\times\cdots\times30109.
\end{equation}
Running the computation over each fixed prime (without parallelization) took $\mathcal{O}$(1 day) using an Apple Silicon M1 Max chip. 
Given the magnitude of this computation, we record $N_{\mathrm{FT}}(f_{10})$ as well as some intermediate counts in \Cref{figure: capacities} and \Cref{table: FT triangle counts}. We find $N_{\mathrm{FT}}(f_{10}) \approx 10^{180}$ which is $70$ orders of magnitude smaller than Anclin's bound. 
Likewise, we applied this algorithm to $f_8$ and $f_9$, resulting in counts of $N_{\mathrm{FT}}(f_8) \approx 10^{35}$ and $N_{\mathrm{FT}}(f_9) \approx 10^{66}$, respectively. See \Cref{table: 2-face upper bounds}. 
Combining the results of \Cref{table: 2-face upper bounds} in (\ref{eq: upper strategy}), using $N_\text{FRT}(f)$ and $N_\text{FT}(f)$ as applicable, gives us: 
\begin{equation}
\label{eq:upper}
    \begin{split}
    N_{\textrm{CY}}^{\Delta^\circ_{491}} &< 1.21 \times 10^{296}.
    \end{split}
\end{equation}

\begin{table}[ht]
\begin{center}
    \begin{tabular}{|p{0.1\linewidth}|p{0.7\linewidth}|p{0.15\linewidth}|}
        \hline
        $n$ & \# fine triangulations of $T_{7,n}$ & magnitude \\
        \hline
        \hline
        7 & \seqsplit{72977202065037} & $7.29 \times 10^{13}$ \\
        \hline
        14 & \seqsplit{67303859310445030532143735447} & $6.73 \times 10^{28}$ \\
        \hline
        21 & \seqsplit{82538241833071662319608077013831474180379171} & $8.25 \times 10^{43}$ \\
        \hline
        28 & \seqsplit{110469482551636815834083564970273245813170066131344202994021} & $1.10 \times 10^{59}$ \\
        \hline
        35 & \seqsplit{153405520601065827395233041403916434565495619834253255337377215694753592286} & $1.53 \times 10^{74}$ \\
        \hline
        42 & \seqsplit{216986756631083658126920483628112951666071397627189154421561640107221284432844801035428489} & $2.16 \times 10^{89}$\\
        \hline
        49 & \seqsplit{310088432970902850840181144241298501449250543248054859810901973445648309893435820168570918245049037410431} & $3.10 \times 10^{104}$ \\
        \hline
        56 & \seqsplit{445892162864331607261451155459356075903267153098914431766555207582747419210216766637190399627372668533451676732583464255} & $4.45 \times 10^{119}$ \\
        \hline
        63 & \seqsplit{643721881546841017370450038311965896835937184278318037273509635473564403389071736713287672464233401686402393603351620144328030764295861} & $6.43 \times 10^{134}$ \\
        \hline
        70 & \seqsplit{931801653317154380405796079127491163178848775602726300987100894320086644389085463600516139586911636316099665808988346144688141670996126976005161734564} & $9.31 \times 10^{149}$\\
        \hline
        77 & \seqsplit{1351307175483914234475846587114225957203398926926719135397039887943607322118917237083168482723058904306581567500091931422178958886851664222447045868021542476069471416} & $1.35 \times 10^{165}$ \\
        \hline
        84 & \seqsplit{1962287132001624119389788641817618767721017439893858857505007108651278332213084991976439732556394117825373997558192408892352073089046072137669318971644281889070086254003018741580611} & $1.96 \times 10^{180}$\\
        \hline
    \end{tabular}
    \caption{The number of fine triangulations of the right lattice triangle defined by the vertices $(0,0),(7,0),$ and $(0,n)$ for $n=7,14,\dots,84$. Note that $T_{7,84}$ corresponds to $f_{10}$.}
    \label{table: FT triangle counts}
\end{center}
\end{table}

\begin{table}[ht]
\centerline{
    \begin{tabular}{ |p{0.07\linewidth}||p{0.35\linewidth}|p{0.1\linewidth}||p{0.35\linewidth}|p{0.1\linewidth}| } 
        \hline
        \textrm{$2$-face} & \textrm{Old Bound (from \cite{Demirtas:2020dbm})} & \textrm{Type} & \textrm{New Bound} & \textrm{Type} \\ 
        \hline
        \hline
        $f_{8}$ & \seqsplit{411376139330301510538742295639337626245683966408394965837152256} & \textrm{Upper FT} & \seqsplit{202115025826361078406926459466100344} & \textrm{Exact FRT} \\
        & \hfill ($\sim 4.11 \times 10^{62}$) & & \hfill ($\sim 2.02\times 10^{35}$) & \\
        \hline
        $f_{9}$ & \seqsplit{17498005798264095394980017816940970922825355447145699491406164851279623993595007385788105416184430592} & \textrm{Upper FT} & \seqsplit{761342982944289349099618507228200078481281500600912757801568059775} & \textrm{Exact FT} \\
        & \hfill ($\sim 1.75 \times 10^{100}$) & & \hfill ($\sim 7.61\times 10^{65}$) & \\
        \hline
        $f_{10}$ & \seqsplit{57277807836949922408837567867349676981443478344341305058882899404622128010705808318690568531649256750858719018437999440148793721514146753400890052083129159241025748615958424204533602522957957552490080016463490494951861107213475167230717574212948590592} & \textrm{Upper FT} & \seqsplit{1962287132001624119389788641817618767721017439893858857505007108651278332213084991976439732556394117825373997558192408892352073089046072137669318971644281889070086254003018741580611} & \textrm{Exact FT} \\
        & \hfill ($\sim 5.73 \times 10^{250}$) & & \hfill ($\sim 1.96\times 10^{180}$) & \\
        \hline
    \end{tabular}
}
\caption{Upper bounds of FRTs of the three remaining 2-faces of our polytope. These constitute our main results on upper bounds.}
\label{table: 2-face upper bounds}
\end{table}

In \Cref{appendix: width-2}, we also provide an alternative recursive formula (\ref{eqn: w2trap}) which suffices for $f_8$. This formula is a generalization of other results in \cite{Kaibel_Ziegler_2003} and is possible because the small width of the face $f_8$ lends itself easily to direct analysis. 

\subsection{Upper bounds for Kreuzer-Skarke}
\begin{figure}[ht]
\centering
	\includegraphics[width=150mm]{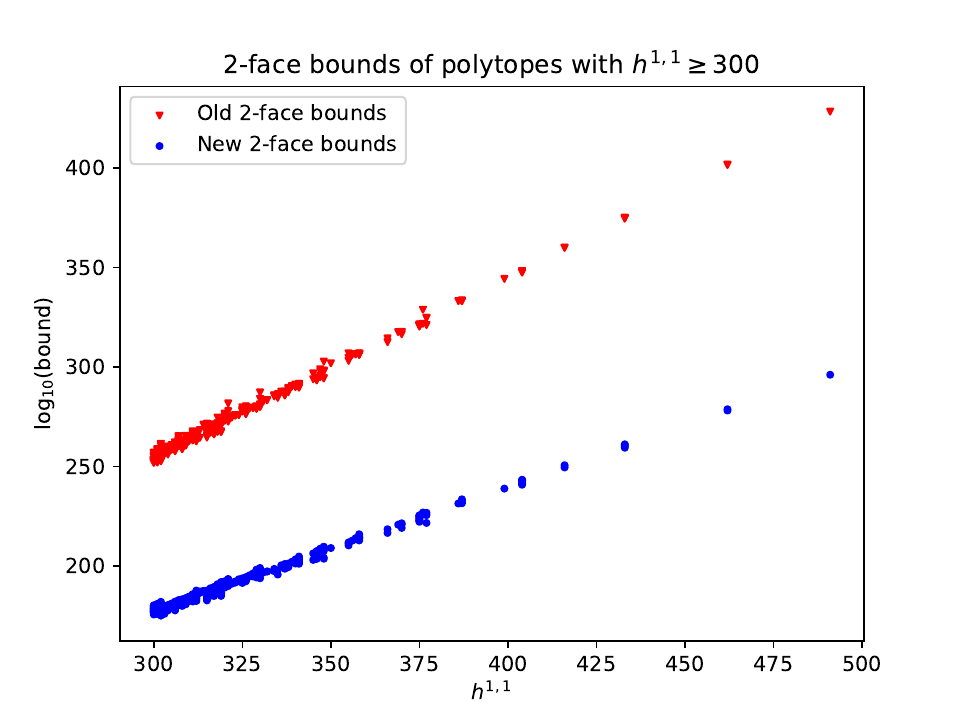} 
	\caption{Improvements in the upper bounds on the number of 2-face equivalence classes for polytopes with $h^{1,1} \geq 300$, comparing those computed in \cite{Demirtas:2020dbm} (red) with those found in this work (blue).}
	\label{fig:improvement}
\end{figure}

We repeat analogous computations and find the number of fine triangulations for the $2$-faces of all 400 other polytopes with $h^{1,1} \geq 300$. Up to affine equivalence, there are an additional 326 such distinct $2$-faces, and all of their counts of fine triangulations 
can be seen in \Cref{appendix: data}. Putting these data together using (\ref{eq: upper strategy}), we find: 

\begin{equation}
\label{eq:upper_300_400}
\sum_{\substack{\text{4D reflexive polytopes } \Delta^\circ \\ \text{with } 300 \leq h^{1,1} < 491}} N_{\text{CY}}^{\Delta^\circ} < 6.89 \times 10^{278}.
\end{equation}
In particular, looking at \Cref{fig:improvement}, we see that the bounds for all polytopes with $h^{1,1} \geq 300$ still appear to approximately follow an exponential trend in $h^{1,1}$, with $\De$ dominating by roughly 20 orders of magnitude.

To determine an upper bound on all Kreuzer-Skarke polytopes, we then make use of the data obtained in \cite{Demirtas:2020dbm}. Specifically, we take their exact upper bound on all of KS (found in footnote 8 of \cite{Demirtas:2020dbm}), and compute the upper bounds as described in their paper for all polytopes with $h^{1,1} \geq 300$. Taking the difference then leaves us with:
\begin{equation}
\label{eq:upper_001_299}
\sum_{\substack{\text{4D reflexive polytopes} \\ \Delta^\circ \text{with }  h^{1,1} < 300}} N_{\text{CY}}^{\Delta^\circ} < 9.33 \times 10^{258}.
\end{equation}
As this is roughly 40 orders of magnitude smaller than (\ref{eq:upper}), we leave it to future work to compute tighter bounds for these polytopes and show that they follow the trend of \Cref{fig:improvement}.

\Cref{eq:upper,eq:upper_300_400,eq:upper_001_299} altogether give our main result:
\begin{equation}
\label{eq:upper_all}
    N_{\textrm{CY}}^{\text{KS}} < 1.211 \times 10^{296}.
\end{equation}

\section{Lower Bounds}\label{section: lower bounds}


\Cref{eq:upper_all} provides an improved upper bound on the count of diffeomorphism classes of CYs. In lieu of an answer to the complementary question of finding a lower bound on the number of such diffeomorphism classes, we work towards answering a related, pragmatic question: that of finding a lower bound on the number of $2$-face equivalence classes. For two triangulations to define non-diffeomorphic CYs, they must have different $2$-face restrictions and hence lie in different classes. However, this count fails to be a lower bound on the count of diffeomorphism classes because the converse need not hold: as in \cite{gendler2023counting, oxford}, distinct 2-face equivalence classes may lie in the same diffeomorphism class. 

Nevertheless, there is utility in such a lower bound as we now explain. No methods are currently known for efficiently checking diffeomorphism at arbitrary $h^{1,1}$ as, most directly, this would require determining if an integral $h^{1,1} \times h^{1,1}$ change of basis matrix exists relating the intersection numbers and second Chern class. Therefore, typically, representatives from $2$-face equivalence classes (commonly called `NTFEs' for non-two-face-equivalent triangulations) are treated as a sufficiently irredundant set of CYs. Moreover, \cite{macfadden2023efficient} demonstrates how to iterate directly over $2$-face equivalence classes efficiently, and so the number of such classes is proportional to the current computational expense of iterating over all potentially-distinct CYs. 


To then bound the number of $2$-face equivalence classes from below, one must understand which collections of triangulations $\{\mathcal{T}_1,\dots, \mathcal{T}_N\}$ of $2$-faces $\{f_1, \dots, f_N\}$ correspond to a $2$-face equivalence class. For a $2$-face equivalence class to exist, there must be some FRST of $\Delta^\circ$ which, upon restriction to each $f_i$, generates the imposed triangulation $\mathcal{T}_i$.
A necessary condition is that each $\mathcal{T}_i$ be fine and regular on $f_i$. However, this is not sufficient -- collections of fine, regular $2$-face triangulations exist which do not admit an FRST of $\Delta^\circ$. This is the notion of `extendability' from \cite{macfadden2023efficient}. Thus, our goal is to bound from below the number of collections of $2$-face triangulations for which such an FRST of $\Delta^\circ$ provably exists.

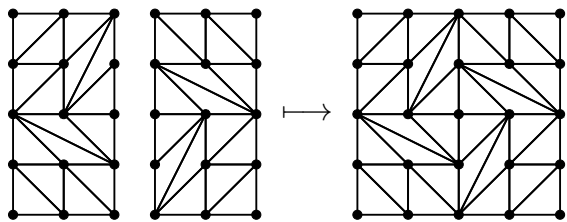
\begin{figure}[htp!]
    \centering
    \begin{minipage}{0.1\textwidth}
    \begin{tikzpicture}
    \tikzmath{ \w = 2; \n = 4;  
    \zero = 0; \one = 1; \two = 2; \three = 3; \four = 4;
    \shift = 0;
    \scale = 1/1.5;
    }

    \foreach \x  [remember=\x as \lastx (initially 0)] in {0,...,\w}{

        \foreach \y [remember=\y as \lasty (initially 0)] in {0,...,\n}{
            \node[draw,circle,inner sep=1.2pt,fill] at (\x*\scale,\y*\scale) {};
        }
    }
    

    \draw [line width=0.25mm] 
    (\zero*\scale,\zero*\scale) --
    (\one*\scale,\zero*\scale) -- (\zero*\scale,\one*\scale) -- cycle;

    \draw [line width=0.25mm] 
    (\one*\scale,\zero*\scale) --
    (\one*\scale,\one*\scale) -- (\zero*\scale,\one*\scale) -- cycle;    

    \draw [line width=0.25mm] 
    (\one*\scale,\one*\scale) --
    (\one*\scale,\zero*\scale) -- (\two*\scale,\zero*\scale) -- cycle;

    \draw [line width=0.25mm] 
    (\two*\scale,\one*\scale) --
    (\two*\scale,\zero*\scale) -- (\one*\scale,\one*\scale) -- cycle;

    \draw [line width=0.25mm] 
    (\zero*\scale,\one*\scale) --
    (\zero*\scale,\two*\scale) -- (\one*\scale,\one*\scale) -- cycle;

    \draw [line width=0.25mm] 
    (\zero*\scale,\two*\scale) --
    (\one*\scale,\one*\scale) -- (\two*\scale,\one*\scale) -- cycle;    

    \draw [line width=0.25mm] 
    (\zero*\scale,\two*\scale) --
    (\two*\scale,\one*\scale) -- (\one*\scale,\two*\scale) -- cycle;

    \draw [line width=0.25mm] 
    (\one*\scale,\two*\scale) --
    (\two*\scale,\two*\scale) -- (\two*\scale,\one*\scale) -- cycle;

    \draw [line width=0.25mm] 
    (\zero*\scale,\two*\scale) --
    (\zero*\scale,\three*\scale) -- (\one*\scale,\three*\scale) -- cycle;

    \draw [line width=0.25mm] 
    (\zero*\scale,\two*\scale) --
    (\one*\scale,\two*\scale) -- (\one*\scale,\three*\scale) -- cycle;    

    \draw [line width=0.25mm] 
    (\zero*\scale,\three*\scale) --
    (\zero*\scale,\four*\scale) -- (\one*\scale,\four*\scale) -- cycle;

    \draw [line width=0.25mm] 
    (\zero*\scale,\three*\scale) --
    (\one*\scale,\three*\scale) -- (\one*\scale,\four*\scale) -- cycle;

    \draw [line width=0.25mm] 
    (\one*\scale,\two*\scale) --
    (\two*\scale,\two*\scale) -- (\two*\scale,\three*\scale) -- cycle;

    \draw [line width=0.25mm] 
    (\one*\scale,\two*\scale) --
    (\two*\scale,\three*\scale) -- (\two*\scale,\four*\scale) -- cycle;    

    \draw [line width=0.25mm] 
    (\one*\scale,\two*\scale) --
    (\two*\scale,\four*\scale) -- (\one*\scale,\three*\scale) -- cycle;

    \draw [line width=0.25mm] 
    (\one*\scale,\three*\scale) --
    (\one*\scale,\four*\scale) -- (\two*\scale,\four*\scale) -- cycle;
    \end{tikzpicture}
    \end{minipage}
    \begin{minipage}{0.02\textwidth}
      
    \end{minipage}
    \begin{minipage}{0.10\textwidth}
    \begin{tikzpicture}
        \tikzmath{ \w = 2; \n = 4;  
        \zero = 0; \one = 1; \two = 2; \three = 3; \four = 4;
        \shift = -2;
        \scale = 1/1.5;
        }

        \foreach \x  [remember=\x as \lastx (initially 0)] in {0,...,\w}{
    
            \foreach \y [remember=\y as \lasty (initially 0)] in {0,...,\n}{
                \node[draw,circle,inner sep=1.2pt,fill] at (\x*\scale,\y*\scale) {};
            }
        }


    \draw [line width=0.25mm] 
    (\two*\scale + \shift*\scale,\zero*\scale) --
    (\three*\scale + \shift*\scale,\zero*\scale) -- (\three*\scale + \shift*\scale,\one*\scale) -- cycle;

    \draw [line width=0.25mm] 
    (\two*\scale + \shift*\scale,\zero*\scale) --
    (\three*\scale + \shift*\scale,\one*\scale) -- (\three*\scale + \shift*\scale,\two*\scale) -- cycle;    

    \draw [line width=0.25mm] 
    (\two*\scale + \shift*\scale,\zero*\scale) --
    (\two*\scale + \shift*\scale,\one*\scale) -- (\three*\scale + \shift*\scale,\two*\scale) -- cycle;

    \draw [line width=0.25mm] 
    (\two*\scale + \shift*\scale,\one*\scale) --
    (\two*\scale + \shift*\scale,\two*\scale) -- (\three*\scale + \shift*\scale,\two*\scale) -- cycle;

    \draw [line width=0.25mm] 
    (\three*\scale + \shift*\scale,\zero*\scale) --
    (\four*\scale + \shift*\scale,\zero*\scale) -- (\four*\scale + \shift*\scale,\one*\scale) -- cycle;

    \draw [line width=0.25mm] 
    (\three*\scale + \shift*\scale,\zero*\scale) --
    (\four*\scale + \shift*\scale,\one*\scale) -- (\three*\scale + \shift*\scale,\one*\scale) -- cycle;    

    \draw [line width=0.25mm] 
    (\three*\scale + \shift*\scale,\one*\scale) --
    (\four*\scale + \shift*\scale,\one*\scale) -- (\four*\scale + \shift*\scale,\two*\scale) -- cycle;

    \draw [line width=0.25mm] 
    (\three*\scale + \shift*\scale,\one*\scale) --
    (\three*\scale + \shift*\scale,\two*\scale) -- (\four*\scale + \shift*\scale,\two*\scale) -- cycle;

    \draw [line width=0.25mm] 
    (\two*\scale + \shift*\scale,\two*\scale) --
    (\two*\scale + \shift*\scale,\three*\scale) -- (\three*\scale + \shift*\scale,\two*\scale) -- cycle;

    \draw [line width=0.25mm] 
    (\two*\scale + \shift*\scale,\three*\scale) --
    (\three*\scale + \shift*\scale,\two*\scale) -- (\four*\scale + \shift*\scale,\two*\scale) -- cycle;    

    \draw [line width=0.25mm] 
    (\two*\scale + \shift*\scale,\three*\scale) --
    (\three*\scale + \shift*\scale,\three*\scale) -- (\four*\scale + \shift*\scale,\two*\scale) -- cycle;

    \draw [line width=0.25mm] 
    (\three*\scale + \shift*\scale,\three*\scale) --
    (\four*\scale + \shift*\scale,\three*\scale) -- (\four*\scale + \shift*\scale,\two*\scale) -- cycle;

    \draw [line width=0.25mm] 
    (\two*\scale + \shift*\scale,\three*\scale) --
    (\two*\scale + \shift*\scale,\four*\scale) -- (\three*\scale + \shift*\scale,\three*\scale) -- cycle;

    \draw [line width=0.25mm] 
    (\two*\scale + \shift*\scale,\four*\scale) --
    (\three*\scale + \shift*\scale,\four*\scale) -- (\three*\scale + \shift*\scale,\three*\scale) -- cycle;    

    \draw [line width=0.25mm] 
    (\three*\scale + \shift*\scale,\three*\scale) --
    (\three*\scale + \shift*\scale,\four*\scale) -- (\four*\scale + \shift*\scale,\three*\scale) -- cycle;

    \draw [line width=0.25mm] 
    (\three*\scale + \shift*\scale,\four*\scale) --
    (\four*\scale + \shift*\scale,\three*\scale) -- (\four*\scale + \shift*\scale,\four*\scale) -- cycle;
    \end{tikzpicture}
    \end{minipage}
    \begin{minipage}{0.05\textwidth}
      $\longmapsto$
    \end{minipage}
    \begin{minipage}{0.17\textwidth}
    \begin{tikzpicture}
    \tikzmath{ \w = 4; \n = 4;  
    \zero = 0; \one = 1; \two = 2; \three = 3; \four = 4;
    \shift = 0;
    \scale = 1/1.5;
    }

    \foreach \x  [remember=\x as \lastx (initially 0)] in {0,...,\w}{

        \foreach \y [remember=\y as \lasty (initially 0)] in {0,...,\n}{
            \node[draw,circle,inner sep=1.2pt,fill] at (\x*\scale,\y*\scale) {};
        }
    }
    

    \draw [line width=0.25mm] 
    (\zero*\scale,\zero*\scale) --
    (\one*\scale,\zero*\scale) -- (\zero*\scale,\one*\scale) -- cycle;

    \draw [line width=0.25mm] 
    (\one*\scale,\zero*\scale) --
    (\one*\scale,\one*\scale) -- (\zero*\scale,\one*\scale) -- cycle;    

    \draw [line width=0.25mm] 
    (\one*\scale,\one*\scale) --
    (\one*\scale,\zero*\scale) -- (\two*\scale,\zero*\scale) -- cycle;

    \draw [line width=0.25mm] 
    (\two*\scale,\one*\scale) --
    (\two*\scale,\zero*\scale) -- (\one*\scale,\one*\scale) -- cycle;

    \draw [line width=0.25mm] 
    (\zero*\scale,\one*\scale) --
    (\zero*\scale,\two*\scale) -- (\one*\scale,\one*\scale) -- cycle;

    \draw [line width=0.25mm] 
    (\zero*\scale,\two*\scale) --
    (\one*\scale,\one*\scale) -- (\two*\scale,\one*\scale) -- cycle;    

    \draw [line width=0.25mm] 
    (\zero*\scale,\two*\scale) --
    (\two*\scale,\one*\scale) -- (\one*\scale,\two*\scale) -- cycle;

    \draw [line width=0.25mm] 
    (\one*\scale,\two*\scale) --
    (\two*\scale,\two*\scale) -- (\two*\scale,\one*\scale) -- cycle;

    \draw [line width=0.25mm] 
    (\zero*\scale,\two*\scale) --
    (\zero*\scale,\three*\scale) -- (\one*\scale,\three*\scale) -- cycle;

    \draw [line width=0.25mm] 
    (\zero*\scale,\two*\scale) --
    (\one*\scale,\two*\scale) -- (\one*\scale,\three*\scale) -- cycle;    

    \draw [line width=0.25mm] 
    (\zero*\scale,\three*\scale) --
    (\zero*\scale,\four*\scale) -- (\one*\scale,\four*\scale) -- cycle;

    \draw [line width=0.25mm] 
    (\zero*\scale,\three*\scale) --
    (\one*\scale,\three*\scale) -- (\one*\scale,\four*\scale) -- cycle;

    \draw [line width=0.25mm] 
    (\one*\scale,\two*\scale) --
    (\two*\scale,\two*\scale) -- (\two*\scale,\three*\scale) -- cycle;

    \draw [line width=0.25mm] 
    (\one*\scale,\two*\scale) --
    (\two*\scale,\three*\scale) -- (\two*\scale,\four*\scale) -- cycle;    

    \draw [line width=0.25mm] 
    (\one*\scale,\two*\scale) --
    (\two*\scale,\four*\scale) -- (\one*\scale,\three*\scale) -- cycle;

    \draw [line width=0.25mm] 
    (\one*\scale,\three*\scale) --
    (\one*\scale,\four*\scale) -- (\two*\scale,\four*\scale) -- cycle;


    \draw [line width=0.25mm] 
    (\two*\scale + \shift*\scale,\zero*\scale) --
    (\three*\scale + \shift*\scale,\zero*\scale) -- (\three*\scale + \shift*\scale,\one*\scale) -- cycle;

    \draw [line width=0.25mm] 
    (\two*\scale + \shift*\scale,\zero*\scale) --
    (\three*\scale + \shift*\scale,\one*\scale) -- (\three*\scale + \shift*\scale,\two*\scale) -- cycle;    

    \draw [line width=0.25mm] 
    (\two*\scale + \shift*\scale,\zero*\scale) --
    (\two*\scale + \shift*\scale,\one*\scale) -- (\three*\scale + \shift*\scale,\two*\scale) -- cycle;

    \draw [line width=0.25mm] 
    (\two*\scale + \shift*\scale,\one*\scale) --
    (\two*\scale + \shift*\scale,\two*\scale) -- (\three*\scale + \shift*\scale,\two*\scale) -- cycle;

    \draw [line width=0.25mm] 
    (\three*\scale + \shift*\scale,\zero*\scale) --
    (\four*\scale + \shift*\scale,\zero*\scale) -- (\four*\scale + \shift*\scale,\one*\scale) -- cycle;

    \draw [line width=0.25mm] 
    (\three*\scale + \shift*\scale,\zero*\scale) --
    (\four*\scale + \shift*\scale,\one*\scale) -- (\three*\scale + \shift*\scale,\one*\scale) -- cycle;    

    \draw [line width=0.25mm] 
    (\three*\scale + \shift*\scale,\one*\scale) --
    (\four*\scale + \shift*\scale,\one*\scale) -- (\four*\scale + \shift*\scale,\two*\scale) -- cycle;

    \draw [line width=0.25mm] 
    (\three*\scale + \shift*\scale,\one*\scale) --
    (\three*\scale + \shift*\scale,\two*\scale) -- (\four*\scale + \shift*\scale,\two*\scale) -- cycle;

    \draw [line width=0.25mm] 
    (\two*\scale + \shift*\scale,\two*\scale) --
    (\two*\scale + \shift*\scale,\three*\scale) -- (\three*\scale + \shift*\scale,\two*\scale) -- cycle;

    \draw [line width=0.25mm] 
    (\two*\scale + \shift*\scale,\three*\scale) --
    (\three*\scale + \shift*\scale,\two*\scale) -- (\four*\scale + \shift*\scale,\two*\scale) -- cycle;    

    \draw [line width=0.25mm] 
    (\two*\scale + \shift*\scale,\three*\scale) --
    (\three*\scale + \shift*\scale,\three*\scale) -- (\four*\scale + \shift*\scale,\two*\scale) -- cycle;

    \draw [line width=0.25mm] 
    (\three*\scale + \shift*\scale,\three*\scale) --
    (\four*\scale + \shift*\scale,\three*\scale) -- (\four*\scale + \shift*\scale,\two*\scale) -- cycle;

    \draw [line width=0.25mm] 
    (\two*\scale + \shift*\scale,\three*\scale) --
    (\two*\scale + \shift*\scale,\four*\scale) -- (\three*\scale + \shift*\scale,\three*\scale) -- cycle;

    \draw [line width=0.25mm] 
    (\two*\scale + \shift*\scale,\four*\scale) --
    (\three*\scale + \shift*\scale,\four*\scale) -- (\three*\scale + \shift*\scale,\three*\scale) -- cycle;    

    \draw [line width=0.25mm] 
    (\three*\scale + \shift*\scale,\three*\scale) --
    (\three*\scale + \shift*\scale,\four*\scale) -- (\four*\scale + \shift*\scale,\three*\scale) -- cycle;

    \draw [line width=0.25mm] 
    (\three*\scale + \shift*\scale,\four*\scale) --
    (\four*\scale + \shift*\scale,\three*\scale) -- (\four*\scale + \shift*\scale,\four*\scale) -- cycle;

    \end{tikzpicture}
    \end{minipage}
    
    \caption{Two fine regular triangulations of $P_{2,4}$ being patched to a single triangulation of $P_{4,4}$. This triangulation is fine but not regular. This example was originally found by Francisco Santos and appears in \cite{Kaibel_Ziegler_2003}.}
    \label{figure: patching}
\end{figure}

To understand how extendability can fail, consider a similar problem of `patching' together two $2$D lattice triangulations\footnote{Although, we stress, this slightly varies from the extendability issue.}. For notation, let $P_{m,n} \coloneqq \{ 0, 1, \dots, m \} \times \{ 0, 1, \dots, n \}$ be the $m\times n$ lattice rectangle. While any fine triangulation of $P_{m,n_1}$ and any other fine triangulation of $P_{m,n_2}$ can be patched together to form a fine triangulation of $P_{m,n_1+n_2}$, the same cannot be said of fine \textit{regular} triangulations. In particular, see \Cref{figure: patching} for an example of two fine, regular triangulations that, when `patched', lead to an irregular triangulation.

This failure mirrors that of extendability: in both cases, despite starting with regular triangulations, there may not be any height vector defining the combined object. There will be some instances in which regularity \textit{can} be preserved via patching which will be discussed in the next subsection.

\subsection{Lower bounds for $\Delta_{491}^\circ$}\label{subsec: f10 lower}
In light of extendability, to generate a lower bound on the number of $2$-face equivalence classes, we must generate \textit{compatible} collections of fine, regular $2$-face triangulations. Given that the upper bound on $\Delta^\circ_{491}$ dominates, we focus our subsequent analysis to this particular polytope. Specifically, we do so by first generating a lower bound on the number of fine, regular triangulations of the largest $2$-face $f_{10}$. From this 2-face, we then lower-bound the number of compatible fine, regular triangulations of the remaining $2$-faces, i.e. those which, together with the triangulation of $f_{10}$, are provably extendable to an FRST of $\De$.

\begin{figure}[ht]
    \centering
    \includegraphics[width=120mm]{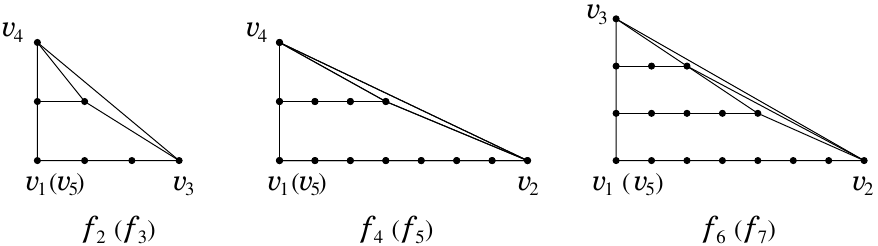}
	\caption{The primary subdivisions of $f_2,\dots,f_7$. Primary subdivisions of $f_8$ and $f_9$ are described in the text. The vertices $v_1,\dots,v_5$ of $\Delta_{491}^\circ$ are numbered as the columns of (\ref{eq:491}).}
    \label{Fig-A}
\end{figure}

As will be shown in \Cref{subsec: lower 491}, any fine regular triangulation of $f_{10}$ is compatible with 
maximal refinements of what we call the \textit{primary subdivisions} of the other $2$-faces. For the primary subdivisions of $f_2,\dots,f_7$, see\footnote{$f_1$ is trivial because it has one triangulation: it is the unit simplex.} \Cref{Fig-A}. Recall from \Cref{table: 2-faces} that $f_8$ is a right triangle of width 2 and height 84; the primary subdivision of $f_8$ is splitting the 2-face into two width 1 strips. Similarly, $f_9$ is a right triangle of width 3 and height 84; the primary subdivision is splitting the $2$-face into three width 1 strips.

We need to both bound, from below, $N_\FRT(f_{10})$, as well as count the number of maximal refinements of the primary subdivisions. First, some technology which motivates the primary subdivisions.

\begin{lemma}[Proposition 3.4 in \cite{Kaibel_Ziegler_2003}]\label{lemma:w2reg}
For $n\geq1$, the following hold:
\begin{gather*}
    N_{\mathrm{FRT}}(P_{1,n}) = N_{\mathrm{FT}}(P_{1,n}) = \binom{2n}{n}, \\
    N_{\mathrm{FRT}}(P_{2,n}) = N_{\mathrm{FT}}(P_{2,n}).
\end{gather*}
\end{lemma}

\Cref{lemma:w2reg} tells us that any fine triangulation of a width $1$ or width $2$ lattice rectangle is regular. We use this extensively with the following lemma, which shows that one can add strips of width $1$ to a regular triangulation without spoiling the regularity.

\begin{lemma}[Lemma 3.5 in \cite{Kaibel_Ziegler_2003}\protect\footnote{As stated, there is a typo in Lemma 3.5 in \cite{Kaibel_Ziegler_2003}. This is however an easy fix: the binomial coefficient is $\binom{2m}{m}$ rather than $\binom{2n}{n}$. The general argument provided still holds true.}]\label{lemma: lower bounding regular} For $m,n \geq 1$, the following holds:
\begin{gather*}
    N_{\mathrm{FRT}}(P_{m+1,n}) \geq  N_{\mathrm{FRT}}(P_{m,n})\cdot \binom{2n}{n}.
\end{gather*}
\end{lemma}

\Cref{lemma: lower bounding regular} can be understood as: given a regular triangulation of $P_{m,n}$, we may append any triangulated width 1 strip and maintain regularity of $P_{m+1,n}$. 
Alternatively, a deconstructive viewpoint may be taken, where we decompose $P_{m,n}$ entirely into width 1 strips, obtaining the lower bound $N_{\mathrm{FRT}}(P_{m,n}) \geq \binom{2n}{n}^m$. This decomposition mirrors our primary subdivisions. Applying the second line of \Cref{lemma:w2reg}, one may slightly improve this bound by decomposing until there is exactly one width 2 strip and a collection of width 1 strips.

\Cref{lemma:w2reg,lemma: lower bounding regular} were proven via a more general lemma regarding regularity of lattice trapezoids of width 1. Note that lattice trapezoids of width 1 with parallel sides of lattice lengths $m$ and $n$ follow a similar formula to \Cref{lemma:w2reg}: the number of fine regular triangulations is $\binom{n+m}{n}$. The upshot of using lattice trapezoids rather than lattice rectangles is that the arguments used in \Cref{lemma:w2reg,lemma: lower bounding regular} directly generalize to the lattice polygons in \Cref{Fig-A}.

\begin{lemma}[Lemma 3.3 in \cite{Kaibel_Ziegler_2003}]\label{lemma: width-1}
    Let $\mathcal{T}$ be a fine triangulation of a lattice trapezoid with two parallel vertical or horizontal sides $S_0$ and $S_1$ at distance one. Every piecewise linear function $h_0 : S_0 \rightarrow \R$ that is strictly convex on $S_0 \cap \Z^2$ can be extended to a lifting function for $\mathcal{T}$. 
\end{lemma}

This lemma provides significant freedom in the choice of lifting function/height vector; for instance, given a lifting function produced by \Cref{lemma: width-1} above, one can produce another lifting function by raising the entire side $S_1$ by a constant height, effectively fixing the height of any single point on $S_1$. We will use this extensively in \Cref{subsec: lower 491}. 
Altogether, these lemmas now guarantee that any fine triangulation refining the primary subdivision of each $f_i$ is also regular. Using the preceding lemmas, we can now count the number of such refinements.

All that remains is to compute a lower bound on $N_\FRT(f_{10})$. Recall from \Cref{table: 2-faces} that $f_{10}$ is a right triangle of width 7 and height 84. We compute a lower bound by decomposing the 2-face into one strip of width 2 and five strips of width 1. Varying the placement of the strip of width 2 and using inclusion-exclusion then gives a lower bound:
\begin{equation}
\label{eq:f10lower}
    \begin{split}
    N_\FRT(f_{10}) &\geq 61629753637666384249294886276486436545694516976935412544055123\\&
    51854904014562566277633806026740716703726964245039569074175193995\\&
    63432314021032777216390798881136475500000\\&
    \approx 6.16 \cdot 10^{167}.
    \end{split}
\end{equation}
One can compute this directly from binomial coefficients and (\ref{eqn: w2trap}).

We claim that any fine regular triangulation of $f_{10}$ and any collection of maximal refinements of the primary subdivisions of $f_2,\dots,f_9$ extends to a fine regular star triangulation of $\De$. We show this below in \Cref{subsec: lower 491}. This implies 
\begin{equation}
\label{eq:lower}
\begin{split}
    N_{\FRST}(\De)
     &\ge N_{\FRT}(f_{10})\cdot\binom{140}  {56}\binom{84}   {28}\cdot\binom{126}  {42}
       \qquad\text{[contribution of $f_{10}$, $f_9$, $f_8$]}\\
     &\qquad\times\binom{11}{4}^2 \binom{6}{2}^2 \cdot\binom{10}{3}^2\cdot\binom{4}{1}^2
    \qquad\text{[contribution of $f_7,\dots,f_2$]}\\
     &\ge 15481007102006939736194364533953509004122317171981255190656\\&
     3311092373914026677314707437577347793417384418452474200793811\\&
     6033667893750406740954155093989036300802356217073401153793835\\&
     5301188418179123818266407383543765321559405609651177459597394\\&
     60369630358603308800000000000000000\\
     &\approx 1.55 \cdot 10^{276},
\end{split}
\end{equation}
where we apply the lower bound (\ref{eq:f10lower}) found above. See \Cref{tab:lower bounds} to compare individual 2-face contributions to either the total number of FRTs of the 2-face (or an upper bound of this quantity); this gives a sense as to how tight the bounds are on individual 2-faces. 

One can look to increase the lower bound by improving the lower bound of $N_{\FRT}(f_{10})$. One way to do this would be to take into account contributions of larger-width trapezoids; this is explored in \Cref{appendix: lower bound improvements}. However, we gain a factor of less than 10 and so omit this from the main text, instead using the bound (\ref{eq:f10lower}) above.

\subsection{Proof of extendability}\label{subsec: lower 491}

Here, we argue that any collection of maximal refinements of the primary subdivisions of the $2$-faces is extendable (i.e., there exists an FRST with said restrictions). 
Let $v_1,\dots,v_5$ be the vertices of $\Delta_{491}^\circ$ numbered
as the columns of (\ref{eq:491}). Then, the $2$-faces are
$$
   f_1=[v_2,v_3,v_4],
$$
$$
   f_2=[v_1,v_3,v_4],\quad
   f_3=[v_5,v_3,v_4],
$$
$$
   f_4=[v_1,v_2,v_4],\quad
   f_5=[v_5,v_2,v_4],
$$
$$
   f_6=[v_1,v_2,v_3],\quad
   f_7=[v_5,v_2,v_3],
$$
$$
   f_8=[v_1,v_4,v_5],\quad
   f_9=[v_1,v_3,v_5],\quad
   f_{10}=[v_1,v_2,v_5],
$$
and the edges have lattice lengths
$$
   |v_1v_5|=84,\quad
   |v_1v_2|=|v_2v_5|=7,\quad
   |v_1v_3|=|v_3v_5|=3,\quad
   |v_1v_4|=|v_4v_5|=2,
$$
$$
   |v_2v_3|=|v_2v_4|=|v_3v_4|=1.
$$

\begin{table}[ht]
    \centering
    \begin{tabular}{|c|c|c|}
         \hline
         2-face & Contribution to the lower bound (\ref{eq:lower}) & Upper bound of $N_{\textrm{FRT}}$ \\
         \hline
         \hline
         $f_1$ & 1 & 1*\\
         \hline
         $f_2,f_3$ & 4 & 5*\\
         \hline
         $f_4,f_5$ & 120 & 204*\\
         \hline 
         $f_6,f_7$ & 4950 & 19594*\\
         \hline
         $f_8$ & $5.09 \times 10^{33}$ & $2.02 \times 10^{35}$*\\
         \hline
         $f_9$ & $8.73 \times 10^{61}$ & $7.61 \times 10^{65}$\\
         \hline
         $f_{10}$ & $6.16 \times 10^{167}$ & $1.96 \times 10^{180}$\\
         \hline
    \end{tabular}
    \caption{Contributions to the lower bound (\ref{eq:lower}) by 2-face. In the right-most column, we also list the total number $N_{\textrm{FT}}$, or, if known, the total number $N_\textrm{FRT}$ of the 2-face. Starred quantities denote exact counts of FRTs.}
    \label{tab:lower bounds}
\end{table}

\if01{
b=Binomial;
N[ b[84+56,56]*b[56+28,28] * b[84+42,42]*(b[7+4,4]*b[4+2,2]*b[7+3,3]*4)^2 ]
}\fi 

Fix an arbitrary collection of triangulations refining the primary subdivisions of $f_i$. We show that there exists an FRST of $\De$ which restricts to this collection of 2-face triangulations. In order to prove this fact, we construct a piecewise linear function $h$ (the height function) on the boundary of $\De$ such that the restriction of $h$ to each 2-face $f_i$ refines the primary subdivisions to the chosen triangulations. We do so by defining height functions $h_i$ on every 2-face which refine the primary subdivisions to triangulations, and which are compatible with one another.

We give a brief overview of the construction of $h_i$: first, we define functions $g_i$ on each 2-face such that lifting face $f_i$ by $g_i$ gives precisely the primary subdivision (not a proper refinement of it). Next, we define functions $k_i$ on each 2-face such that the restriction of $k_i$ to any \textit{cell} of the primary subdivision of $f_i$ agrees with the triangulation on this cell; note that lifting the entire 2-face $f_i$ by $k_i$ need not give the chosen triangulation of $f_i$. However, when adding $k_i$ to a sufficiently large multiple of $g_i$, then $k_i$ acts as a perturbation of $g_i$, and lifting $f_i$ by this sum gives our triangulation. Note that care still must be taken in the construction to ensure compatibility between 2-faces, as we see below.

\begin{figure}[ht]
\centering
	\includegraphics[width=150mm]{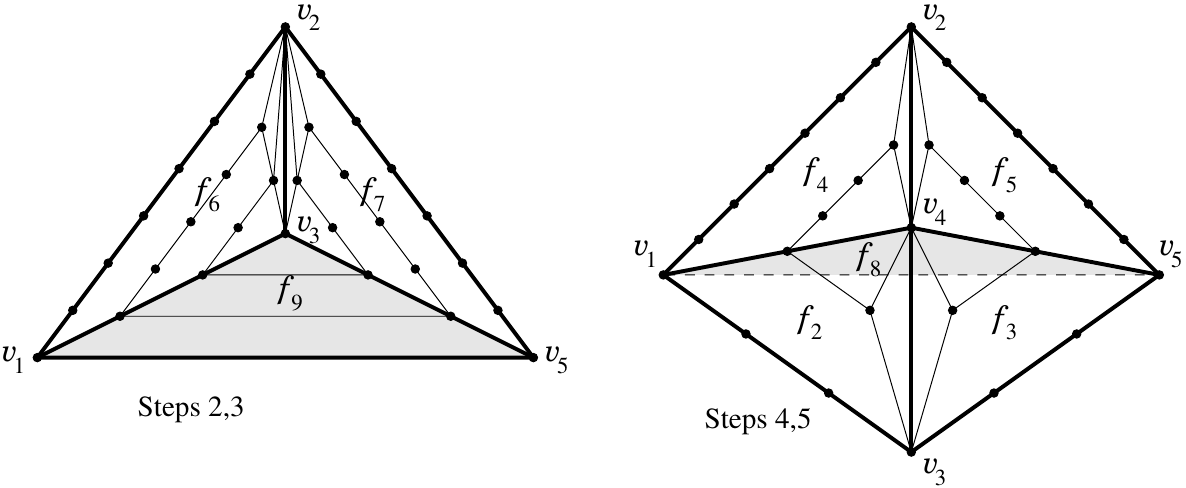} 
	\caption{Left: a diagram of the $3$-face $1235$ of $\De$. The face $f_{10}$ is obscured. Dots indicate lattice points, except for $f_9$ which has $172$ lattice points, so we just shade it gray. The displayed lines indicate the primary subdivisions, as in \Cref{Fig-A}. Right: an analogous diagram showing details of the rest of polytope $\De$, now with the 3-face $1235$ as the base. The remaining faces with $v_4$ as a vertex are visible. Face $f_8$ cuts through the figure and is represented by the shaded region above the dashed line.}
	\label{Fig-lb}
\end{figure}

The construction of $h$ goes as follows (see Figure~\ref{Fig-lb}).

\smallskip
Step 1. By hypothesis we may define such an $h$ on $f_{10}$. By \Cref{lemma: width-1} the restriction of $h$ to the edge $[v_1,v_5]$ can be extended (strip by strip) to piecewise linear functions $k_8,k_9$ on $f_8,f_9$ such that on any cell of the respective 2-face, $k_8$ and $k_9$ refine the cell to the imposed triangulation.

\smallskip
Step 2. Using \Cref{lemma: width-1} we extend (strip by strip) the restriction $h|_{[v_1,v_2]}$ from $f_{10}$ to a piecewise linear function $k_6$ on $f_6=[v_1,v_2,v_3]$ so that $k_6$ is compatible with the chosen imposed triangulation of each cell of the primary subdivision and so that the restriction of $k_6$ to the edge $[v_1,v_3]$ coincides with $k_9$. We may impose this latter restriction because of the freedom afforded to us by the lemma. We also extend $h|_{[v_5,v_2]}$ to a function $k_7$ on $f_7=[v_5,v_2,v_3]$ in the same way, in particular, so that $k_7|_{[v_5,v_3]}=k_9|_{[v_5,v_3]}$.

\smallskip
Step 3. Let $g_6,g_7,g_9$ be piecewise linear functions on $f_6,f_7,f_9$ compatible with the primary
subdivisions and such that
\begin{enumerate}
    \item $g_6|_{[v_1,v_3]}=g_9|_{[v_1,v_3]}$,
    \item $g_7|_{[v_5,v_3]}=g_9|_{[v_5,v_3]}$, and
    \item $g_6|_{[v_1,v_2]} = g_7|_{[v_5,v_2]} = g_9|_{[v_1,v_5]} = 0$.
\end{enumerate}
Then we define $h_i = k_i + Ag_i$ on the 2-faces $f_i$, $i=6,7,9$ for $A\gg 1$. This now ensures that the various $h_i$ generate the desired triangulations on $f_i$, and that the $h_i$ are compatible on their shared faces. We now define $h$ on $f_i$, $i=6,7,9$ to be $h_i$. Therefore, we have defined $h$ on the boundary of the $3$-face $[v_1,v_2,v_3,v_5]$. 

\smallskip
Step 4. (Similar to Step 2.)
Using \Cref{lemma: width-1} we extend $h$ 
as follows:
\begin{enumerate}
    \item from $[v_1,v_3]$ to $k_2$ on $f_2 = [v_1,v_3,v_4]$,
    \item from $[v_5,v_3]$ to $k_3$ on $f_3 = [v_5,v_3,v_4]$,
    \item from $[v_1,v_2]$ to $k_4$ on $f_4 = [v_1,v_2,v_4],$ and
    \item from $[v_5,v_2]$ to $k_5$ on $f_5 = [v_5,v_2,v_4]$.
\end{enumerate}
As above, we perform these extensions so that the resulting functions $k_2,\dots, k_5$
are piecewise linear, compatible with the triangulations of the cells of the primary subdivisions, and
so that
$k_2|_{[v_1,v_4]}=k_4|_{[v_1,v_4]}=k_8|_{[v_1,v_4]}$ and
$k_3|_{[v_5,v_4]}=k_5|_{[v_5,v_4]}=k_8|_{[v_5,v_4]}$.

\smallskip
Step 5. (Similar to Step 3.)
Let $g_2,\dots,g_5$, and $g_8$ be the piecewise linear functions on $f_2,\dots,f_5$ which are compatible
with the primary subdivisions, vanish on the edges where $h$ is already defined
(i.e. on the edges of the 3-face $[v_1,v_2,v_3,v_5]$), and coincide on the common edges; thus
these functions take the same value at $v_4$ and they take the same value at the midpoints
of the segments $[v_1,v_4]$ and $[v_5,v_4]$.
We define $h_i$ on $f_i$, $i=2,3,4,5,8$, to be $h_i = k_i + B g_i$ with $B\gg A$.

\smallskip
Step 6. Finally, we define $h$ on 2-faces $f_i$, $i=2,3,4,5,8$ to be $h_i$. We extend $h$ to $f_1$ by linearity. We further impose that the height of the origin is sufficiently small such that $h$ also defines a star triangulation. Now, the constructed $h$ defines the FRST of interest.

\section{Conclusion}\label{section: conclusion}

In this work, we have demonstrated that Batyrev's construction applied to FRSTs of $4$D reflexive polytopes defines at most $10^{296}$ diffeomorphism classes of Calabi-Yau threefolds, improving \cite{Demirtas:2020dbm}'s bound of $10^{428}$. Both bounds arise by bounding the number of $2$-face equivalence classes as in (\ref{eq: upper strategy}) -- the only difference is that for $2$-faces with more than 17 lattice points, we used the exact number of fine triangulations while \cite{Demirtas:2020dbm} used Anclin's upper bound on the number of fine triangulations.

Specifically, we studied the number of $2$-face equivalence classes for polytopes with $h^{1,1}\geq300$. We studied only polytopes with $h^{1,1}\geq300$ since the pre-existing bounds from \cite{Demirtas:2020dbm} depend roughly exponentially on $h^{1,1}$, with all $h^{1,1}<300$ polytopes contributing only $10^{259}$ to the bound. We find that this exponential dependence is still true after our improvement (see \Cref{fig:improvement}); namely, our final bound is still ultimately set by the largest-$h^{1,1}$ polytope, $\Delta^\circ_{491}$. Indeed, we show that the polytopes with $300\leq h^{1,1}<491$ contribute $10^{279}$ to the upper bound, orders-of-magnitude smaller than $\Delta^\circ_{491}$'s contribution of $10^{296}$.

In addition to the above results on upper bounds of the number of $2$-face equivalence classes, we demonstrated a lower bound of $10^{276}$ such classes for $\Delta^\circ_{491}$. This was obtained by defining `primary subdivisions' of the faces $f_1, \dots, f_9$, analogous to the decompositions utilized in \cite{Kaibel_Ziegler_2003}, and proving that, for any FRT of $f_{10}$ and any maximal refinement of said subdivisions, there exists an FRST of $\Delta^\circ_{491}$ with said $2$-face restrictions. That is, these $2$-face triangulations are extendable. Thus, our upper bound of $10^{296}$ $2$-face equivalence classes is loose by at most $20$ orders of magnitude. Of particular interest for this upper bound are $f_9$ and $f_{10}$, for which we utilized counts $N_\FT$ in (\ref{eq: upper strategy}). If one instead had exact counts $N_\FRT$ for these faces, this upper bound could drop by at most $17$ orders of magnitude (see \Cref{tab:lower bounds}).

The state-of-the-art in generating potentially-inequivalent CYs is by constructing representatives of each $2$-face equivalence class as in \cite{macfadden2023efficient}. Thus, using current methods, studies of all Kreuzer-Skarke CYs with $h^{1,1}=491$ must operate on at \textit{least} $10^{276}$ CYs. It is computationally infeasible to study $10^{276}$ CYs -- even if such a hypothetical study took $1$ femtosecond per CY, it would still take at least $10^{261}$ seconds to study all of them. For such a study to be viable, one would need an algorithm to directly construct (not stepping through, e.g., $2$-face equivalence classes) an exponentially smaller collection of all potentially non-diffeomorphic CYs, should it exist.

One natural direction for future study is to 
adapt these bounds in light of the generalizations to the construction of CYs from $4$D reflexive polytopes discussed in \cite{vex,Berglund_2018}. Currently, this is non-trivial as it requires defining a clean notion of 2-face equivalence for such CYs. Another direction would be to better determine the true distribution of inequivalent CYs. A first step towards this goal would be to obtain a non-trivial lower bound on the number of diffeomorphism classes of CYs, as opposed to the lower bound on 2-face equivalence classes found in this work. To the best of the authors' knowledge, such a non-trivial lower bound on diffeomorphism classes is not known; moreover, despite $\Delta^\circ_{491}$ having at least $10^{276}$ 2-face equivalence classes, it has not been demonstrated that these define at least two non-diffeomorphic CYs. One approach could be coming up with novel invariants of the Wall data; while we can compute the triple intersection numbers and second Chern class, it is hard to determine whether or not there exists a change of basis between these data. On the other hand, the invariants in \cite{gendler2023counting,oxford} (such as the GCD invariants) are easy to work with but have low discriminative power and are hence well-suited only to low-$h^{1,1}$ CYs. Searching for more powerful but still computationally practical invariants is a worthwhile endeavor.

\section*{Acknowledgments}

NM and MS are grateful to Volker Kaibel, Liam McAllister, Andres Rios-Tascon, and Francisco Santos for helpful correspondence. MS would also like to thank Pyry Kuusela for providing valuable feedback on an earlier version of the draft. The work of NM and MS is supported in part by NSF grant PHY-2309456.

\appendix
\section{Right lattice trapezoids of width $2$}
\label{appendix: width-2}

As discussed in \Cref{section: upper bounds}, for the largest $2$-faces $f_8$, $f_9$, and $f_{10}$ of $\De$, we use a recursive algorithm of \cite{orevkov2022counting} to compute the exact count of fine triangulations. However, following \cite{Kaibel_Ziegler_2003}, we also introduce a recursive algorithm designed specifically to tackle right lattice trapezoids of width 2, including the third-largest $2$-face $f_8$. This algorithm additionally can be used when computing the bound (\ref{eq:f10lower}).

First setting notation following \cite{Kaibel_Ziegler_2003,orevkov2022counting}, define
\begin{align}\label{eq: notation}
    f(m,n) \coloneqq N_{\text{FT}}(P_{m,n}) & \quad & \text{and} & \quad & f^{\mathrm{reg}}(m,n) \coloneqq N_{\text{FRT}}(P_{m,n}).
\end{align}

Section 2.1 of \cite{Kaibel_Ziegler_2003} discusses determining the exact count of fine triangulations of narrow lattice strips via a recursive algorithm. For lattice trapezoids of width $m=1$, with parallel sides of lattice lengths $a$ and $b$, the number of fine triangulations is exactly $g_1(a,b) \coloneqq \binom{a+b}{a} = \binom{a+b}{b}$. 

For strips of width $m=2$ and arbitrary height $n$, \cite{Kaibel_Ziegler_2003} makes the observation that to count all fine triangulations, one can enumerate the triangulations by their highest \textit{width 2 diagonal}, where a width 2 diagonal is an internal edge of horizontal length two but lattice length 1 (i.e., its midpoint is not a lattice point). This diagonal then decomposes the width 2 strip into two `non-interacting' width 1 strips which lie above the diagonal, as well as the trapezoid determined by the diagonal. See \Cref{figure:lattice2} below.

This observation still holds when considering \textit{right-lattice trapezoids} of width $m=2$. We consider those whose upper boundary has integral slope (which describes e.g. $f_8$ and the width 2 strips used in \Cref{subsec: f10 lower}'s bound of $N_{\FRT}(f_{10})$), but note that this can be further generalized. Let our lattice trapezoid be determined by $\conv\{(0,0),(0,n),(2,0),(2,n+2k)\}$ where $k$ is a non-negative integer. In analog with (\ref{eq: notation}), let $f(2;n,k)$ (resp. $f^{\textrm{reg}}(2;n,k)$) denote the number of fine (resp. fine regular) triangulations of this trapezoid. 

Then, to count all fine triangulations, we count the number of fine triangulations for each configuration determined by a placement of the highest width 2 diagonal $\overline{AB}$. This can be split into two terms: (1) the lack of a placement of a diagonal, and (2) a valid placement of the upper diagonal $\overline{AB}$, where $A$ lies on the left boundary of the trapezoid and $B$ on the right boundary. Term (1) can be computed as the product of the number of fine triangulations of the non-interacting width 1 strips.  Each summand in term (2), determined by $\overline{AB}$, can be computed as the product of the number of fine triangulations of the non-interacting strips above the diagonal and the number of fine triangulations of the trapezoid $\conv\{(0,0),(0,A),(2,0),(2,B)\}$. Again, see \Cref{figure:lattice2} below for this decomposition. Let $g_2(A,B)$ be the number of fine triangulations of this trapezoid, where the explicit recursion for $g_2(A,B)$ can be found in \cite{Kaibel_Ziegler_2003}. Then:

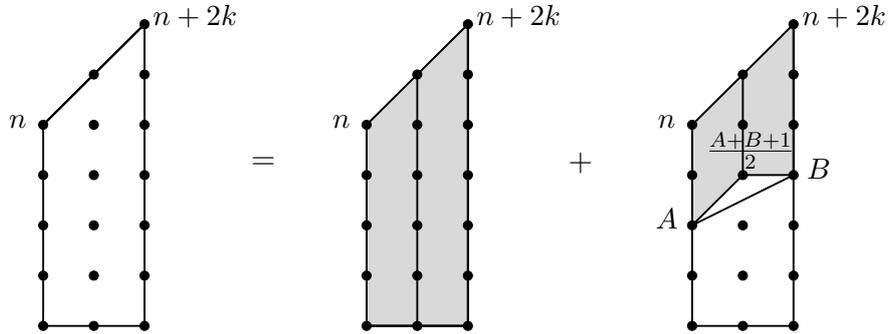
\begin{figure}[htp!]
  \centering

  \begin{minipage}{0.2\textwidth}
  \begin{tikzpicture}

    \tikzmath{ \w = 2; \n = 4; \k = 1;  
    \xA = 0; \xMid = 1; \xB = 2; 
    \yA = 4; \yMid = 5; \yB = 6; 
    \lHeight = \n + \xA*\k; \rHeight = \n + \xB*\k; 
    \scale = 1/1.5;
    }

    \coordinate (A)   at (\xA*\scale,\yA*\scale);
    \coordinate (B)   at (\xB*\scale,\yB*\scale);
    \coordinate (mdpt)   at (\xMid*\scale,\yMid*\scale);

    \filldraw[fill=lightgray!60]((\xA*\scale,\yA*\scale) -- (\xMid*\scale,\yMid*\scale) -- (\xB*\scale,\yB*\scale) -- (\xB*\scale,\rHeight*\scale) -- (\xA*\scale,\lHeight*\scale) -- cycle;);

    \foreach \x  [remember=\x as \lastx (initially 0)] in {0,...,\w}{
        
        \pgfmathsetmacro{\z}{\n + \x*\k}
        \ifthenelse{\x > 0}{ \pgfmathsetmacro{\lastz}{\n  + (\x-1)*\k}}{}

        \foreach \y [remember=\y as \lasty (initially 0)] in {0,...,\z}{
            \node[draw,circle,inner sep=1.2pt,fill] at (\x*\scale,\y*\scale) {};

            \ifthenelse{\x = 0 \AND \y > 0}{\draw [line width=0.25mm] (\x*\scale,\y*\scale) -- (\x*\scale,\lasty*\scale);}{}

            \ifthenelse{\x = \w \AND \y > 0}{\draw [line width=0.25mm] (\x*\scale,\y*\scale) -- (\x*\scale,\lasty*\scale);}{}

            \ifthenelse{\y = 0}{\draw [line width=0.25mm] (\x*\scale,\y*\scale) -- (\lastx*\scale,\lasty*\scale);}{}

            \ifthenelse{\x > 0 \AND \y = \z}{\draw [line width=0.25mm] (\x*\scale,\y*\scale) -- (\lastx*\scale,\lastz*\scale);}{}
        }
    }

    \draw [line width=0.25mm] (A) -- (B) -- (mdpt) -- cycle;

    \pgfmathsetmacro{\z}{\n + \xA*\k};
    \foreach \y in {\yA,...,\z}{
        \ifthenelse{\y < \z}{\pgfmathsetmacro{\h}{\y + 1};}{\pgfmathsetmacro{\h}{\y};}
        \draw [line width=0.25mm] (\xA*\scale,\y*\scale) -- (\xA*\scale,\h*\scale);
    }

    \pgfmathsetmacro{\z}{\n + \xMid*\k};
    \foreach \y in {\yMid,...,\z}{
        \ifthenelse{\y < \z}{\pgfmathsetmacro{\h}{\y + 1};}{\pgfmathsetmacro{\h}{\y};}
        \draw [line width=0.25mm] (\xMid*\scale,\y*\scale) -- (\xMid*\scale,\h*\scale);
    }

    \pgfmathsetmacro{\z}{\n + \xB*\k};
    \foreach \y in {\yB,...,\z}{
        \ifthenelse{\y < \z}{\pgfmathsetmacro{\h}{\y + 1};}{\pgfmathsetmacro{\h}{\y};}
        \draw [line width=0.25mm] (\xB*\scale,\y*\scale) -- (\xB*\scale,\h*\scale);
    }

    \node[label={[shift = {(-0.5*\scale,-0.5*\scale)}]\small $n$}] at (0,\n*\scale) {};
    \node[label={[shift = {(+1.0*\scale,-0.5*\scale)}]\small $n+\w k$}] at (\w*\scale,\n*\scale+\w*\k*\scale) {};

  \end{tikzpicture}
  \end{minipage}
    \begin{minipage}{0.05\textwidth}
        $=$
    \end{minipage}
  \begin{minipage}{0.2\textwidth}
  \begin{tikzpicture}

    \tikzmath{ \w = 2; \n = 4; \k = 1;  
    \xA = 0; \xMid = 1; \xB = 2; 
    \yA = 0; \yMid = 0; \yB = 0; 
    \lHeight = \n + \xA*\k; \rHeight = \n + \xB*\k; 
    \scale = 1/1.5;
    }

    \coordinate (A)   at (\xA*\scale,\yA*\scale);
    \coordinate (B)   at (\xB*\scale,\yB*\scale);
    \coordinate (mdpt)   at (\xMid*\scale,\yMid*\scale);

    \filldraw[fill=lightgray!60]((\xA*\scale,\yA*\scale) -- (\xMid*\scale,\yMid*\scale) -- (\xB*\scale,\yB*\scale) -- (\xB*\scale,\rHeight*\scale) -- (\xA*\scale,\lHeight*\scale) -- cycle;);

    \foreach \x  [remember=\x as \lastx (initially 0)] in {0,...,\w}{
        
        \pgfmathsetmacro{\z}{\n + \x*\k}
        \ifthenelse{\x > 0}{ \pgfmathsetmacro{\lastz}{\n  + (\x-1)*\k}}{}

        \foreach \y [remember=\y as \lasty (initially 0)] in {0,...,\z}{
            \node[draw,circle,inner sep=1.2pt,fill] at (\x*\scale,\y*\scale) {};

            \ifthenelse{\x = 0 \AND \y > 0}{\draw [line width=0.25mm] (\x*\scale,\y*\scale) -- (\x*\scale,\lasty*\scale);}{}

            \ifthenelse{\x = \w \AND \y > 0}{\draw [line width=0.25mm] (\x*\scale,\y*\scale) -- (\x*\scale,\lasty*\scale);}{}

            \ifthenelse{\y = 0}{\draw [line width=0.25mm] (\x*\scale,\y*\scale) -- (\lastx*\scale,\lasty*\scale);}{}

            \ifthenelse{\x > 0 \AND \y = \z}{\draw [line width=0.25mm] (\x*\scale,\y*\scale) -- (\lastx*\scale,\lastz*\scale);}{}
        }
    }

    \draw [line width=0.25mm] (A) -- (B) -- (mdpt) -- cycle;

    \pgfmathsetmacro{\z}{\n + \xA*\k};
    \foreach \y in {\yA,...,\z}{
        \ifthenelse{\y < \z}{\pgfmathsetmacro{\h}{\y + 1};}{\pgfmathsetmacro{\h}{\y};}
        \draw [line width=0.25mm] (\xA*\scale,\y*\scale) -- (\xA*\scale,\h*\scale);
    }

    \pgfmathsetmacro{\z}{\n + \xMid*\k};
    \foreach \y in {\yMid,...,\z}{
        \ifthenelse{\y < \z}{\pgfmathsetmacro{\h}{\y + 1};}{\pgfmathsetmacro{\h}{\y};}
        \draw [line width=0.25mm] (\xMid*\scale,\y*\scale) -- (\xMid*\scale,\h*\scale);
    }

    \pgfmathsetmacro{\z}{\n + \xB*\k};
    \foreach \y in {\yB,...,\z}{
        \ifthenelse{\y < \z}{\pgfmathsetmacro{\h}{\y + 1};}{\pgfmathsetmacro{\h}{\y};}
        \draw [line width=0.25mm] (\xB*\scale,\y*\scale) -- (\xB*\scale,\h*\scale);
    }

    \node[label={[shift = {(-0.5*\scale,-0.5*\scale)}]\small $n$}] at (0,\n*\scale) {};
    \node[label={[shift = {(+1.0*\scale,-0.5*\scale)}]\small $n+\w k$}] at (\w*\scale,\n*\scale+\w*\k*\scale) {};

  \end{tikzpicture}
  \end{minipage}
  \begin{minipage}{.05\textwidth}
      +
  \end{minipage}
  \begin{minipage}{0.2\textwidth}
  \begin{tikzpicture}

    \tikzmath{ \w = 2; \n = 4; \k = 1;  
    \xA = 0; \xMid = 1; \xB = 2; 
    \yA = 2; \yMid = 3; \yB = 3; 
    \lHeight = \n + \xA*\k; \rHeight = \n + \xB*\k; 
    \scale = 1/1.5;
    }

    \coordinate (A)   at (\xA*\scale,\yA*\scale);
    \coordinate (B)   at (\xB*\scale,\yB*\scale);
    \coordinate (mdpt)   at (\xMid*\scale,\yMid*\scale);

    \filldraw[fill=lightgray!60]((\xA*\scale,\yA*\scale) -- (\xMid*\scale,\yMid*\scale) -- (\xB*\scale,\yB*\scale) -- (\xB*\scale,\rHeight*\scale) -- (\xA*\scale,\lHeight*\scale) -- cycle;);

    \foreach \x  [remember=\x as \lastx (initially 0)] in {0,...,\w}{
        
        \pgfmathsetmacro{\z}{\n + \x*\k}
        \ifthenelse{\x > 0}{ \pgfmathsetmacro{\lastz}{\n  + (\x-1)*\k}}{}

        \foreach \y [remember=\y as \lasty (initially 0)] in {0,...,\z}{
            \node[draw,circle,inner sep=1.2pt,fill] at (\x*\scale,\y*\scale) {};

            \ifthenelse{\x = 0 \AND \y > 0}{\draw [line width=0.25mm] (\x*\scale,\y*\scale) -- (\x*\scale,\lasty*\scale);}{}

            \ifthenelse{\x = \w \AND \y > 0}{\draw [line width=0.25mm] (\x*\scale,\y*\scale) -- (\x*\scale,\lasty*\scale);}{}

            \ifthenelse{\y = 0}{\draw [line width=0.25mm] (\x*\scale,\y*\scale) -- (\lastx*\scale,\lasty*\scale);}{}

            \ifthenelse{\x > 0 \AND \y = \z}{\draw [line width=0.25mm] (\x*\scale,\y*\scale) -- (\lastx*\scale,\lastz*\scale);}{}
        }
    }

    \draw [line width=0.25mm] (A) -- (B) -- (mdpt) -- cycle;

    \pgfmathsetmacro{\z}{\n + \xA*\k};
    \foreach \y in {\yA,...,\z}{
        \ifthenelse{\y < \z}{\pgfmathsetmacro{\h}{\y + 1};}{\pgfmathsetmacro{\h}{\y};}
        \draw [line width=0.25mm] (\xA*\scale,\y*\scale) -- (\xA*\scale,\h*\scale);
    }

    \pgfmathsetmacro{\z}{\n + \xMid*\k};
    \foreach \y in {\yMid,...,\z}{
        \ifthenelse{\y < \z}{\pgfmathsetmacro{\h}{\y + 1};}{\pgfmathsetmacro{\h}{\y};}
        \draw [line width=0.25mm] (\xMid*\scale,\y*\scale) -- (\xMid*\scale,\h*\scale);
    }

    \pgfmathsetmacro{\z}{\n + \xB*\k};
    \foreach \y in {\yB,...,\z}{
        \ifthenelse{\y < \z}{\pgfmathsetmacro{\h}{\y + 1};}{\pgfmathsetmacro{\h}{\y};}
        \draw [line width=0.25mm] (\xB*\scale,\y*\scale) -- (\xB*\scale,\h*\scale);
    }

    \node[label={[shift = {(-0.5*\scale,-0.5*\scale)}]\small $n$}] at (0,\n*\scale) {};
    \node[label={[shift = {(+1.0*\scale,-0.5*\scale)}]\small $n+\w k$}] at (\w*\scale,\n*\scale+\w*\k*\scale) {};
    \node[label={[shift = {(-0.5*\scale,-0.5*\scale)}]\small $A$}] at (\xA*\scale,\yA*\scale) {};
    \node[label={[shift = {(+0.5*\scale,-0.5*\scale)}]\small $B$}] at (\xB*\scale,\yB*\scale) {};
    \node[label={[shift = {(0.15*\scale,-0.3*\scale)}]\small $\frac{A+B+1}{2}$}] at (\xMid*\scale,\yMid*\scale) {};

  \end{tikzpicture}
  \end{minipage}
  
  \caption{The deconstruction of fine lattice triangulations into two non-interacting width 1 strips and a sum over triangulations determined by the highest width 2 diagonal $\overline{AB}$. The sum over diagonals is left implicit.} 
  \label{figure:lattice2}
\end{figure}

\begin{equation}\label{eqn: w2trap}
    \begin{aligned}
        f(2;n,k) &= \binom{2n+k}{n}\binom{2n+3k}{n+2k} \\
        & + \sum_{\substack{0 \leq A \leq n \\ 0 \leq B \leq n+2k \\ A + B \equiv 1 \, (\mod 2)}} g_2(A,B)\binom{2n+k-\frac{3A+B+1}{2}}{n-A}\binom{2n+3k-\frac{A+3B+1}{2}}{n+2k-B}.
    \end{aligned}
\end{equation}

As this lattice trapezoid is width $2$, we note that as in \Cref{lemma:w2reg}: 
\begin{equation}
    f^{\textrm{reg}}(2;n,k) = f(2;n,k).
\end{equation}
Therefore, we have a recursive algorithm which can be used to find the exact number of fine, regular triangulations of right lattice trapezoids whose upper boundary is integral. Extending this style of formula beyond width 2 lattice trapezoids above can be done, but will be less efficient than the method described in detail in \Cref{appendix: method}. 

\section{Counting fine triangulation of rectangles of a fixed width}
\label{appendix: method}
The improvement \cite{orevkov2022counting} of the dynamic programming algorithm from \cite{Kaibel_Ziegler_2003}
is as follows.

An {\it admissible shape} $S$ of width $m$ is a lattice polygon which is cut from
the strip $[0,m]\times[0,+\infty]$ by the graph of a non-negative continuous piecewise-linear function $\phi$, i.e.,
$$
   S = \{(x,y)\mid 0\le x\le m,\; 0\le y\le \phi(x)\}.
$$
We say that a lattice polygon $Q$ is a {\it primitive tile} in the following three cases:
\begin{enumerate}
\item $Q$ is a primitive lattice triangle without vertical sides;
\item $Q$ is a primitive lattice triangle whose vertical side is contained
      in the boundary of the strip $0 \le x \le m$;
\item $Q = \Delta_1 \cup \Delta_2$ where $\Delta_1$ and $\Delta_2$ are primitive lattice triangles
such that $\Delta_1 \cap \Delta_2$ is a common vertical side of $\Delta_1$ and $\Delta_2$.
\end{enumerate}
A primitive tile $Q$ is {\it $S$-maximal} for an admissible shape $S$ if $Q\subset S$ and the upper part of the
boundary of $Q$ is contained in the upper part of the boundary of $S$, i.e.,
$$
   \max_{(x,y)\in Q} y = \max_{(x,y)\in S} y
$$
for any $x$ such that $Q\cap(\{x\}\times\mathbb R)$ is non-empty.

We say that $S'$ is an {\it admissible subshape} of an admissible shape $S$ and we write $S'\prec S$ if $S'$
is the closure of $S\setminus(Q_1\cup\dots\cup Q_k)$, $k\ge 1$, where $Q_1,\dots,Q_k$ are $S$-maximal primitive
tiles with pairwise disjoint interiors. In this case we set $\#(S',S)=k$.
Note that the relation “to be an admissible subshape” is not transitive.

Denote the number of primitive triangulations of an admissible shape $S$ by $f(S)$.
The following lemma is the inclusion-exclusion formula in our setting (see \cite[Lemma 2.2]{orevkov2022counting};
cf.~\cite[Lemma 2.2]{Kaibel_Ziegler_2003}).

\begin{figure}
\centering
		\includegraphics[width=100mm]{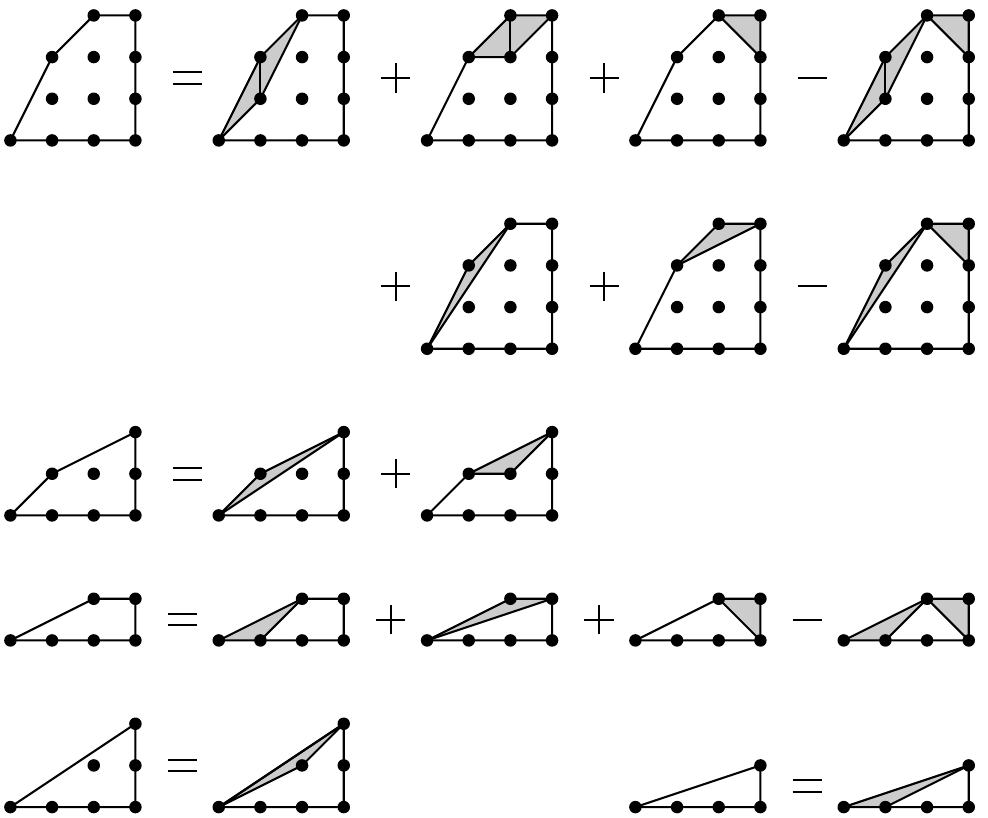} 
	\caption{Inclusion-exclusion formula for admissible shapes.}
	\label{fig.algo}
\end{figure}

\begin{lemma}\label{lem.inclu-exclu}
For any admissible shape $S$ we have (see Figure~\ref{fig.algo})
\begin{equation}\label{eq.inclu-exclu}
   f(S) = \sum_{S'\prec S} (-1)^{\#(S',S)-1} f(S').
\end{equation}
\end{lemma}

This gives us the following dynamic programming algorithm for computation of $f(S)$ for
a given shape $S$ of width $m$. We start by setting $f(S)=1$ for $S=[0,m]$ (the shape
with the empty interior).
Then, successively for $A=1,2,\dots,2mn$, we compute $f(S)$ for all shapes (included in $P_{m,n}$)
of area $A$ by formula \eqref{eq.inclu-exclu}, using the previously computed shapes of smaller areas
(here we normalize the area so that the area of primitive lattice triangles is $1$).
The area of a shape with the upper boundary $[(x_0,y_0),(x_1,y_1),\dots,(x_k,y_k)]$ ($x_0=0$, $x_k=m$)
is equal to
\begin{equation}\label{eq.area}
   \sum_{i=0}^k s_iy_i, \qquad\text{where $s_i=(x_{i+1}-x_{i-1})$ and $x_{-1}=0$, $x_{k+1}=m$}.
\end{equation}

The following is a {\tt Wolfram Mathematica} code that computes the number of primitive triangulations
of the triangles $T_{3,3n}$, $n=1,\dots,28$ (recall that $f_9\cong T_{3,84}$).
The computation takes $\sim$2 seconds.

\begin{verbatim}
F[0,0,0]=1;            (* F[b,c,d] = f( [(0,0),(1,b),(2,c),(3,d),(3,0)] ) *)
Do[                          (* G1[c,d] = f( [(0,0), (2,c),(3,d),(3,0)] ) *)
   Do[ d = area - 2*b - 2*c; (* G2[b,d] = f( [(0,0),(1,b), (3,d),(3,0)] ) *)
     If[ d<0 || d>84, Continue[], F[b,c,d]=0 ];
     If[ b>0, F[b,c,d] += F[b-1,c,d] ];  If[ c>0, F[b,c,d] += F[b,c-1,d] ];
     If[ d>0, F[b,c,d] += F[b,c,d-1]; If[ b>0, F[b,c,d] -= F[b-1,c,d-1] ] ];
     If[ 2*c==b+d+1, F[b,c,d] += G2[b,d] ];
     If[ 2*b==c+1, F[b,c,d] += G1[c,d]; If[ d>0, F[b,c,d] -= G1[c,d-1] ] ],
   {b,0,28},{c,0,56}];
   Do[ d = area - 3*c; b = (c-1)/2;
     If[ d<0 || d>84, Continue[], G1[c,d] = F[b,c,d] ];
     If[ d>0, G1[c,d] += (G1[c,d-1] - F[b,c,d-1]) ];
     If[ 3*c == 2*d + 1, G1[c,d] += G2[b,d] ],
   {c,1,55,2}];
   Do[ d = (area - 3*b)/2; c = (b+d-1)/2;
     If[ d<0 || d>84 || Mod[b+d,2]==0, Continue[], G2[b,d] = F[b,c,d] ];
     If[ 3*b == d + 1, G2[b,d] += G1[c,d] ],
   {b,Mod[area,2],28,2}];
   If[ Mod[area,9]==0, b=area/9; Print[F[b,2b,3b]] ],
{area,1,3*84}]
\end{verbatim}

\section{Estimates of the complexity}
\label{appendix: complexity}
Here we discuss the complexity of the above algorithm in the case of rectangles $P_{m,n}=[0,m]\times[0,n]$.
The total number of admissible shapes is bounded above by $(n+2)^{m+1}$. Indeed, each shape $S\subset P_{m,n}$
can be encoded by a sequence $(y_0,\dots,y_m)$, where $0\le y_k\le m$ if $(k,y_k)$ belongs to the upper
boundary of $S$, and $y_k=n+1$ if the upper boundary does not contain an integral point on the vertical line $x=k$. 

The arguments as in the proof of \cite[Lemma 2.5]{Kaibel_Ziegler_2003} give a (rather coarse) upper bound $3\cdot 2^{m-1}$ for the
number of admissible subshapes of a given shape. Thus the total number of the additions in \eqref{eq.inclu-exclu}
during the whole computation is bounded by
$$
     3\cdot 2^{m-1}(n+2)^{m+1}.
$$
However, as it is pointed out in \cite{Kaibel_Ziegler_2003}, ``the bottleneck in the computations is always memory'',
and this is still so 20 years after. So, let us discuss the needed memory in more details.

When computing $f(S)$ for the shapes of a given area $A$, we need in fact to keep in the memory only the values
of $f(S')$ for the shapes $S'$ such that $A-m\le\text{Area}(S')\le A$. Let us estimate the number of such shapes.
It is convenient to do it in probabilistic terms.
Let $N_A$  be the number of shapes of area $A$.
For a given subset $\xi=\{x_0,\dots,x_k\}\subset\{0,\dots,m\}$ such that $x_0=0$ and $x_k=m$, 
let $N_A(\xi)$ be the number of shapes $S$ of area $A$ contained in $P_{m,n}$ such that the integral points on the upper boundary of $S$
are $(x_i,y_i)$ for some integers $y_0,\dots,y_k$. Let $s_0,\dots,s_k$ be as in \eqref{eq.area}.
Then we have
$$
     N_A(\xi) = \#\big\{ (y_0,\dots,y_k)\in\mathbb Z^{k+1} \;\big|\; s_0 y_0 + \dots + s_k y_k = A, \; 0\le y_i\le n\big\}.
$$
Let $X_0,\dots,X_k$ be independent random variables such that each $X_i$ takes the values in $\{0,\dots,n\}$
with equal probabilities.
Then $N_A(\xi)= (n+1)^{k+1} \proba(s_0 X_0+\dots+s_k X_k=A)$.
We have $s_0,s_k\ge 1$ and $s_i\ge 2$ if $0<i<k$. Hence $s_0^2\ge 2s_0-1$, $s_k^2\ge 2s_k-1$, and $s_i^2\ge 2s_i$ if $0<i<k$.
We also have $\sum s_i=2m$. Therefore: 
$$
   s_0^2 + \dots + s_k^2 \ge 2(s_0+\dots+s_k)-2 = 4m-2.
$$
Since the variance of $X_i$ is $((n+1)^2-1)/12$,
by the Central Limit Theorem we have
$$
     N_A(\xi)\approx   (n+1)^k \sqrt{\frac{12}{2\pi(s_0^2+\dots+s_k^2)}}
     \le (n+1)^k \sqrt{\frac{3}{\pi(2m-1)}}
      < \frac{(n+1)^k}{\sqrt{2m-1}}
$$
and then 
$$
     N_A= \sum_k \sum_{|\xi|=k+1} N_A(\xi)
      \lesssim \sum_k \binom{m-1}{k-1}\frac{(n+1)^k}{\sqrt{2m-1}}
      < \frac{(n+2)^m}{\sqrt{2m-1}}.
$$
Thus the number of the values of $f(S)$ that we need to keep simultaneously in the memory is bounded above by
the quantity 
$$
   \frac{N_{A-m}+\dots+N_A}2 \lesssim \frac{(m+1)(n+2)^m}{2\sqrt{2m-1}}
           < \frac{(n+2)^m\sqrt{2m+7}}4
$$
(the denominator 2 in the last formula comes from the symmetry of the rectangle $P_{m,n}$
with respect to the vertical axis).

By the same arguments we obtain that the maximal number of values of $f(S)$
that we need to keep in the memory during the
computation of $f(T_{m,n})$ is 
$$
    C_m n^m + O(n^{m-1}), \quad\text{where}\quad
    C_m = \frac{(m+1)!}{m^m \sqrt{2\pi \left(\frac{m^3}9 - \frac{m^2}{12} + \frac{m}{18}\right)}}
        = \big(3+\tfrac{35}{8m}+O(m^{-2})\big)e^{-m}.
$$ 

\section{Further possible improvements to the lower bound of $N_{\FRT}(f_{10})$}
\label{appendix: lower bound improvements}
\begin{figure}[h]
\centering
		\includegraphics[width=80mm]{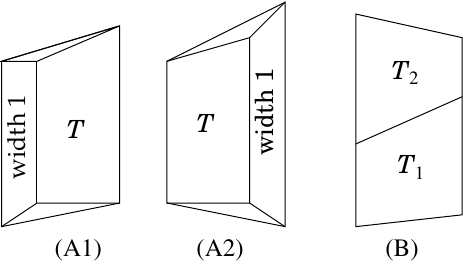} 
	\caption{Recursive construction of FRTs. Any vertical side of $T$
	         (and even both of them) may have zero length, i.e. $T$ may be degenerated to a triangle or a segment.
	         One of the vertical sides (but not both) of $T_i$ also may have zero length.}
	\label{figT}
\end{figure}

There are further possible improvements that one can make to the lower bound of $N_\FRT(f_{10})$.
One can recursively construct FRTs by the following distinct operations.
Given FRTs of some trapezoids $T$, $T_1$, $T_2$, we can obtain an FRT of the larger
trapezoids as shown in Figure \ref{figT}, where 
the common edge of $T_1$ and $T_2$
has lattice length 1. 
The width 1 patterns should be chosen so that the triangles in Figure \ref{figT} are
minimal possible (in particular each of the triangle should be replaced by a segment if possible).
It is not difficult to compute recursively the number of all FRTs of $f_{10}$ that are obtained
in this way. 

\begin{figure}[h]
\centering
		\includegraphics[width=120mm]{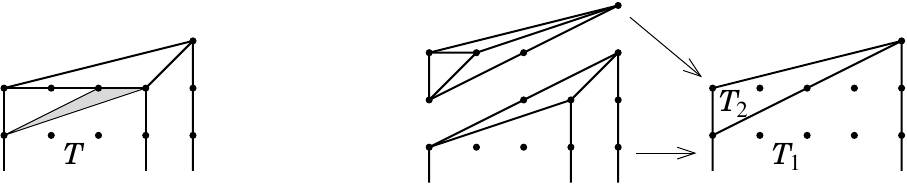} 
	\caption{Precaution when the common edges in the operation (B) are allowed to have lattice length 2.
	      If $T$ is involved in the operation (A2) and its triangulation contains the gray triangle,
	      then this triangulation can be also obtained by the operation (B), where the triangle $T_2$
	      is obtained by the operation (A1) from a segment (considered a trapezoid whose both
	      vertical sides have zero length).
	      }
	\label{figT4}
\end{figure}

This construction can be generalized by allowing the common edge of $T_1$ and $T_2$ (see Figure \ref{figT})
to have lattice length 1 or 2. In this case the computation becomes more complicated because we have to
exclude repeated counting of some FRTs; see Figure~\ref{figT4}.

Some other improvements of this kind are also possible. For example, one can generalize the operations in
Figure~\ref{figT}(B) to pentagons and hexagons, but the control of repeated counting becomes more and more complicated with less and less gain. The cumulative effect of all the improvements discussed in this section is an increase in the lower bound by a factor of less than 10. We thus do not explain them in detail.

\section{Triangulation data for all relevant $2$-faces}
\label{appendix: data}
We first give in \Cref{table: 2-face vertices and multiplicities} a list of the $2$-faces $\{f\}$ (up to affine equivalence) appearing in Kreuzer-Skarke polytopes with $h^{1,1} \geq 300$ as well as their associated fine triangulation counts $N_\textrm{FT}(f)$. The list of $2$-faces is indexed by: (1) area, from least to greatest, (2) number of edges, from least to greatest, and (3) edge lengths, treated lexicographically. In addition to an index, we present a list of vertices defining the $2$-face, and note which polytopes the $2$-face appears in. The $401$ polytopes with $h^{1,1} \geq 300$ are 1-indexed, and are presented in order of increasing $h^{1,1}$. For polytopes with the same $h^{1,1}$, they are ordered in the same manner as is found on a webpage maintained by Skarke \cite{skarke_webpage}. We then denote the multiplicity of a given $2$-face appearing in a polytope by an exponent. 

\Cref{table: all 2-face counts} then follows the indexing in \Cref{table: 2-face vertices and multiplicities} and provides an exact count of $N_{\text{FT}}(f)$. We note that for all $2$-faces with at most 17 lattice points in this list, $N_{\text{FT}}(f) = N_{\text{FRT}}(f)$, except for the $2$-faces 70, 76, and 77, which have $N_{\text{FRT}}(f)$ equal to 14295, 19594, and 37085 respectively. These data (along with those computed in \cite{Demirtas:2020dbm}) can then be used to reproduce the upper bounds (\ref{eq:upper}) and (\ref{eq:upper_300_400}). They can also be downloaded from \cite{paper_data}.

As described in \Cref{appendix: method}, the dynamical programming algorithm used in this paper to compute the number of fine triangulations for a lattice polygon also computes as intermediate data the number of fine triangulations of all admissible shapes (of the same width and with the same lower boundary) of the polygon. We note that while \Cref{appendix: method} talks about the special case in which the lower boundary is the horizontal segment $[0,m] \times \{0\}$, we may also allow the lower boundary to be the graph of a continuous piecewise-linear function $\phi_0$ such that $\phi_0(x) \leq \phi(x)$ for all $x \in [0,m]$ (see \cite[Section~2.1]{orevkov2022counting}). All of the 2-faces described below of width greater than three can be embedded in this more general way into one of the following 19 lattice polygons: $157, 185, 228, 237, 242, 244, 256, 287, 291, 293, 310, 314, 316, 323, 325, 328, 330, 332$ or $333$. This style of computation was shown, for instance, in \Cref{table: FT triangle counts}.



\newpage
{\small
\newgeometry{left=1cm,right=1cm,top=2cm,bottom=1in}
\begin{longtable}{|p{0.5cm}|p{7.2cm}|p{10.2cm}|}
\caption{$2$-faces and their appearences in KS polytopes with $h^{1,1} \geq 300$.}\label{table: 2-face vertices and multiplicities}\\
\hline
Ind. & $2$-face Vertices & $\text{Polytope}^{\text{Multiplicity}}$ \\
\hline
\hline
\endfirsthead
\hline
Ind. & $2$-face Vertices & $\text{Polytope}^{\text{Multiplicity}}$ \\
\hline
\hline
\endhead
\hline
1 & (0, 0), (0, 1), (1, 0) & \(1^{4}\), \(2^{2}\), \(3^{3}\), \(4^{4}\), \(5^{4}\), \(6^{3}\), \(7^{3}\), \(8^{2}\), \(9^{4}\), \(10^{4}\), \(11^{3}\), \(12^{3}\), \(13^{4}\), \(14^{3}\), \(15^{3}\), \(16^{5}\), \(17^{7}\), \(18^{8}\), \(19^{14}\), \(20^{5}\), \(21^{4}\), \(22^{6}\), \(23^{3}\), \(24^{4}\), \(25^{1}\), \(26^{4}\), \(27^{6}\), \(28^{3}\), \(29^{5}\), \(30^{4}\), \(31^{2}\), \(32^{1}\), \(33^{4}\), \(34^{7}\), \(35^{10}\), \(36^{15}\), \(37^{6}\), \(38^{3}\), \(39^{4}\), \(40^{6}\), \(41^{7}\), \(42^{4}\), \(43^{4}\), \(44^{2}\), \(45^{5}\), \(46^{14}\), \(47^{12}\), \(48^{2}\), \(49^{5}\), \(50^{2}\), \(51^{6}\), \(52^{3}\), \(53^{2}\), \(54^{5}\), \(55^{5}\), \(56^{8}\), \(57^{2}\), \(58^{3}\), \(59^{6}\), \(60^{7}\), \(61^{7}\), \(62^{12}\), \(63^{1}\), \(64^{2}\), \(65^{2}\), \(66^{2}\), \(67^{3}\), \(68^{5}\), \(69^{2}\), \(70^{3}\), \(71^{5}\), \(72^{4}\), \(73^{8}\), \(74^{5}\), \(75^{6}\), \(76^{3}\), \(77^{2}\), \(78^{7}\), \(79^{2}\), \(80^{2}\), \(81^{5}\), \(82^{5}\), \(83^{4}\), \(84^{7}\), \(85^{4}\), \(86^{6}\), \(87^{8}\), \(88^{4}\), \(89^{1}\), \(90^{2}\), \(91^{6}\), \(92^{5}\), \(93^{8}\), \(94^{7}\), \(95^{11}\), \(96^{2}\), \(97^{4}\), \(98^{5}\), \(99^{3}\), \(100^{11}\), \(101^{14}\), \(102^{6}\), \(103^{7}\), \(104^{10}\), \(105^{2}\), \(106^{1}\), \(107^{1}\), \(108^{3}\), \(109^{3}\), \(110^{4}\), \(111^{2}\), \(112^{5}\), \(113^{5}\), \(114^{1}\), \(115^{2}\), \(116^{4}\), \(117^{10}\), \(118^{11}\), \(119^{13}\), \(120^{11}\), \(121^{8}\), \(122^{3}\), \(123^{3}\), \(124^{5}\), \(125^{2}\), \(126^{2}\), \(127^{4}\), \(128^{7}\), \(129^{11}\), \(130^{9}\), \(131^{12}\), \(132^{2}\), \(133^{2}\), \(134^{5}\), \(135^{5}\), \(136^{4}\), \(137^{7}\), \(138^{6}\), \(139^{9}\), \(140^{13}\), \(141^{2}\), \(142^{2}\), \(143^{5}\), \(144^{8}\), \(145^{11}\), \(146^{2}\), \(147^{9}\), \(148^{2}\), \(149^{2}\), \(150^{5}\), \(151^{3}\), \(152^{5}\), \(153^{3}\), \(154^{5}\), \(155^{2}\), \(156^{3}\), \(157^{6}\), \(158^{5}\), \(159^{6}\), \(160^{9}\), \(161^{4}\), \(162^{6}\), \(163^{10}\), \(164^{12}\), \(165^{3}\), \(166^{2}\), \(167^{2}\), \(168^{2}\), \(169^{3}\), \(170^{3}\), \(171^{5}\), \(172^{5}\), \(173^{3}\), \(174^{5}\), \(175^{5}\), \(176^{6}\), \(177^{5}\), \(178^{12}\), \(179^{2}\), \(180^{4}\), \(181^{2}\), \(182^{2}\), \(183^{2}\), \(184^{2}\), \(185^{5}\), \(186^{2}\), \(187^{3}\), \(188^{6}\), \(189^{7}\), \(190^{9}\), \(191^{7}\), \(192^{6}\), \(193^{2}\), \(194^{5}\), \(195^{6}\), \(196^{9}\), \(197^{10}\), \(198^{6}\), \(199^{1}\), \(200^{2}\), \(201^{1}\), \(202^{1}\), \(203^{1}\), \(204^{4}\), \(205^{2}\), \(206^{2}\), \(207^{2}\), \(208^{2}\), \(209^{4}\), \(210^{4}\), \(211^{2}\), \(212^{2}\), \(213^{2}\), \(214^{2}\), \(215^{12}\), \(216^{3}\), \(217^{1}\), \(218^{2}\), \(219^{1}\), \(220^{3}\), \(221^{1}\), \(222^{4}\), \(223^{3}\), \(224^{12}\), \(225^{15}\), \(226^{6}\), \(227^{1}\), \(228^{3}\), \(229^{2}\), \(230^{4}\), \(231^{4}\), \(232^{4}\), \(233^{4}\), \(234^{11}\), \(235^{12}\), \(236^{14}\), \(237^{4}\), \(238^{2}\), \(239^{2}\), \(240^{1}\), \(241^{4}\), \(242^{2}\), \(244^{1}\), \(245^{5}\), \(246^{2}\), \(247^{5}\), \(248^{8}\), \(249^{6}\), \(250^{5}\), \(251^{2}\), \(252^{9}\), \(253^{5}\), \(254^{3}\), \(255^{11}\), \(256^{2}\), \(257^{1}\), \(258^{4}\), \(259^{2}\), \(260^{2}\), \(261^{5}\), \(262^{2}\), \(263^{3}\), \(264^{6}\), \(265^{9}\), \(266^{12}\), \(267^{2}\), \(268^{5}\), \(269^{2}\), \(270^{3}\), \(271^{5}\), \(272^{3}\), \(273^{4}\), \(274^{4}\), \(275^{3}\), \(276^{2}\), \(277^{4}\), \(278^{3}\), \(279^{4}\), \(280^{3}\), \(281^{5}\), \(282^{6}\), \(283^{4}\), \(284^{4}\), \(285^{2}\), \(286^{1}\), \(287^{7}\), \(288^{2}\), \(289^{2}\), \(290^{5}\), \(291^{2}\), \(292^{2}\), \(293^{4}\), \(294^{1}\), \(295^{5}\), \(296^{11}\), \(297^{1}\), \(298^{4}\), \(299^{10}\), \(300^{13}\), \(301^{4}\), \(302^{9}\), \(303^{2}\), \(304^{2}\), \(305^{4}\), \(306^{2}\), \(307^{8}\), \(308^{2}\), \(309^{2}\), \(310^{5}\), \(311^{2}\), \(312^{3}\), \(313^{6}\), \(314^{2}\), \(315^{2}\), \(316^{2}\), \(317^{5}\), \(318^{3}\), \(319^{5}\), \(320^{3}\), \(321^{5}\), \(322^{2}\), \(323^{6}\), \(324^{1}\), \(325^{2}\), \(326^{1}\), \(327^{1}\), \(328^{4}\), \(329^{2}\), \(330^{2}\), \(331^{2}\), \(332^{4}\), \(333^{4}\), \(334^{2}\), \(335^{2}\), \(336^{4}\), \(337^{1}\), \(338^{3}\), \(339^{12}\), \(340^{6}\), \(341^{1}\), \(342^{4}\), \(343^{4}\), \(344^{4}\), \(345^{11}\), \(346^{14}\), \(347^{2}\), \(348^{2}\), \(349^{3}\), \(350^{9}\), \(351^{2}\), \(352^{2}\), \(353^{5}\), \(354^{3}\), \(355^{4}\), \(356^{4}\), \(357^{2}\), \(358^{4}\), \(359^{3}\), \(360^{5}\), \(361^{1}\), \(362^{10}\), \(363^{2}\), \(364^{2}\), \(365^{2}\), \(366^{5}\), \(367^{1}\), \(368^{2}\), \(369^{1}\), \(370^{1}\), \(371^{4}\), \(372^{2}\), \(373^{2}\), \(374^{4}\), \(375^{2}\), \(376^{1}\), \(377^{3}\), \(378^{1}\), \(379^{4}\), \(380^{11}\), \(381^{2}\), \(382^{4}\), \(383^{4}\), \(384^{2}\), \(385^{5}\), \(386^{2}\), \(387^{1}\), \(388^{2}\), \(389^{1}\), \(390^{4}\), \(391^{2}\), \(392^{4}\), \(393^{4}\), \(394^{2}\), \(395^{1}\), \(396^{2}\), \(397^{1}\), \(398^{4}\), \(399^{1}\), \(400^{4}\), \(401^{1}\) \\
\hline
\end{longtable}
\restoregeometry
}
{\small
\newgeometry{left=1cm,right=1cm,top=2cm,bottom=1in}
\captionof{table}{$2$-faces and their associated fine triangulation counts.}\label{table: all 2-face counts}
\centering
\begin{minipage}{0.45\textwidth}
\begin{tabular}{|c|p{6.7cm}|}
\hline
Index & $N_\text{FT}(f)$ \\
\hline
\hline
1 & 1 \\ \hline
2 & 1 \\ \hline
3 & 2 \\ \hline
4 & 1 \\ \hline
5 & 1 \\ \hline
6 & 3 \\ \hline
7 & 1 \\ \hline
8 & 1 \\ \hline
9 & 4 \\ \hline
10 & 2 \\ \hline
11 & 4 \\ \hline
12 & 6 \\ \hline
13 & 1 \\ \hline
14 & 4 \\ \hline
15 & 5 \\ \hline
16 & 10 \\ \hline
17 & 1 \\ \hline
18 & 1 \\ \hline
19 & 5 \\ \hline
20 & 6 \\ \hline
21 & 8 \\ \hline
22 & 15 \\ \hline
23 & 1 \\ \hline
24 & 5 \\ \hline
25 & 7 \\ \hline
26 & 18 \\ \hline
27 & 21 \\ \hline
28 & 24 \\ \hline
29 & 10 \\ \hline
30 & 13 \\ \hline
31 & 28 \\ \hline
32 & 53 \\ \hline
33 & 79 \\ \hline
34 & 36 \\ \hline
35 & 84 \\ \hline
36 & 1 \\ \hline
37 & 31 \\ \hline
38 & 28 \\ \hline
39 & 71 \\ \hline
40 & 98 \\ \hline
41 & 120 \\ \hline
42 & 210 \\ \hline
43 & 1 \\ \hline
44 & 11 \\ \hline
45 & 117 \\ \hline
46 & 330 \\ \hline
47 & 300 \\ \hline
\end{tabular}
\end{minipage}
\begin{minipage}{0.45\textwidth}
\begin{tabular}{|c|p{6.7cm}|}
\hline
Index & $N_\text{FT}(f)$ \\
\hline
\hline
48 & 1 \\ \hline
49 & 104 \\ \hline
50 & 156 \\ \hline
51 & 454 \\ \hline
52 & 734 \\ \hline
53 & 13 \\ \hline
54 & 522 \\ \hline
55 & 204 \\ \hline
56 & 378 \\ \hline
57 & 609 \\ \hline
58 & 91 \\ \hline
59 & 1576 \\ \hline
60 & 717 \\ \hline
61 & 786 \\ \hline
62 & 4713 \\ \hline
63 & 2574 \\ \hline
64 & 1048 \\ \hline
65 & 2933 \\ \hline
66 & 3782 \\ \hline
67 & 7549 \\ \hline
68 & 2183 \\ \hline
69 & 9570 \\ \hline
70 & 14303 \\ \hline 
71 & 7273 \\ \hline
72 & 10660 \\ \hline
73 & 21368 \\ \hline
74 & 47256 \\ \hline
75 & 7164 \\ \hline
76 & 19611 \\ \hline 
77 & 37097 \\ \hline 
78 & 101547 \\ \hline
79 & 141671 \\ \hline
80 & 49488 \\ \hline
81 & 1 \\ \hline
82 & 1 \\ \hline
83 & 668517487 \\ \hline
84 & 41634159376 \\ \hline
85 & 294384302208 \\ \hline
86 & 13322475161785816128 \\ \hline
87 & 94592566195943835168 \\ \hline
88 & 126069338181924507924 \\ \hline
89 & 503803582032732726348 \\ \hline
90 & 671738109376976968464 \\ \hline
91 & 12166488474600849404472 \\ \hline
92 & 895307096508249363876 \\ \hline
93 & 3578214261159145784376 \\ \hline
94 & 4770952348212194379168 \\ \hline
\end{tabular}
\end{minipage}
\begin{longtable*}{|c|p{15cm}|}
\hline
Index & $N_\text{FT}(f)$ \\
\hline
\hline
\endfirsthead
\hline
Index & $N_\text{FT}(f)$ \\
\hline
\hline
\endhead
\hline
95 & 89695519146920262088884 \\ \hline
96 & 6359080291811668807348 \\ \hline
97 & 33889469562467768628048 \\ \hline
98 & 240753501292063537182144 \\ \hline
99 & 320915259993885481804920 \\ \hline
100 & 1282874862704479693368840 \\ \hline
101 & 1710499816939306257825120 \\ \hline
102 & 35689780867280180104306320 \\
\hline \end{longtable*}
\restoregeometry
}

\bibliographystyle{utphys}
\bibliography{master}

\end{document}